\documentclass[11pt]{article}
\usepackage{epsf}
\usepackage{./chicago} 
\usepackage{./jasaalan} 
\usepackage[letterpaper, top=3cm, bottom=3cm]{geometry}
\usepackage[latin1]{inputenc}
\usepackage{setspace}
\usepackage{fancyhdr}
\usepackage{tocloft}
\usepackage{epsfig}

\usepackage{lscape}
\usepackage[normalem]{ulem}

\makeatletter
\newcommand{\al}{\alpha}
\newcommand{\be}{\beta}
\newcommand{\tht}{\theta}

\newcommand{\la}{\lambda}
\newcommand{\La}{\Lambda}
\newcommand{\de}{\delta}

\newcommand{\ga}{\gamma}

\newcommand{\delim}[3]{\left#1#2\right#3}

\newcommand{\bm}[1]{\mbox{\boldmath$#1$}}

\newcommand{\ep}{\epsilon}

\newcommand{\std}{\sigma}
\newcommand{\var}{\std^{2}}

\newcommand{\er}{Erd\H{o}s-R\'{e}nyi }

\newcommand{\ws}{Watts-Strogatz }

\begin{document}

\title{Topics in Social Network Analysis \\
and Network Science}
\author{A. James O'Malley\footnote{Department of Health Care Policy, Harvard Medical School, Boston, MA 02115. Email: omalley@hcp.med.harvard.edu} \and Jukka-Pekka Onnela\footnote{Department of Biostatistics, Harvard School of Public Health, Boston, MA 02115. Email: onnela@hsph.harvard.edu}}

\maketitle

\begin{abstract}
This chapter introduces statistical methods used in the analysis of social networks and in the rapidly evolving parallel-field of network science. Although several instances of social network analysis in health services research have appeared recently, the majority involve only the most basic methods and thus scratch the surface of what might be accomplished. Cutting-edge methods using relevant examples and illustrations in health services research are provided.
\end{abstract}


\section*{Part I: Introduction and Background}
\addtocounter{section}{1}

Social network analysis is the study of the structure of relationships linking individuals (or other social units, such as organizations) and of interdependencies in behavior or attitudes related to configurations of social relations. The observational units in a social network are the relationships between individuals and their attributes. Whereas studies in medicine typically involve individuals whose observations can be thought of as statistically independent, observations made on social networks may be simultaneously dependent on all other observations due to the social ties and pathways linking them. Accordingly, different statistical techniques are needed to analyze social network data. The focus of this chapter is {\it sociocentric data}, the case when relational data is available for all pairs of individuals, allowing a fully-fledged review of available methods. 

Two major questions in social network analysis are: 1) do behavioral and other mutable traits spread from person-to-person through a process of induction (also known as {\it social influence}, {\it peer effects}, or {\it social contagion}); 2) what exogeneous factors (e.g., shared actor traits) or endogeneous factors (e.g., internal configurations of actors such as triads) are important to the overall structure of relationships among a group of individuals. 

The first problem has affinity to medical studies in that individuals are the observational units. In medicine, the health of an individual is paramount and so individual outcomes have historically been used to judge the effectiveness of an intervention. A study of social influence in medicine may involve the same outcome but the treatment or intervention is the same variable evaluated on the peers of the focal individual (referred to as {\it alters}). An important characteristic of studies of social influence is that individuals may partly or fully share treatments and one individual's treatment may depend on the outcome of another. For example, an intervention that encourages person A to exercise in order to lose weight might also influence the weight of A's friends (B and C) because they exercise more when around A. Hence, A's weight intervention may also affect the weight of B and C. A consequence is that the total effect of A's treatment must also consider it's effect on B and C, the benefit to individuals to whom B and C are connected, and so on. Such {\it interference} between observations violates the stable-unit treatment value assumption (SUTVA) that one individuals treatment not affect anothers outcome \cite{Rubi:1978}, which presents challenges for identification of causal effects. Interference is likely to result in an incongruity between a regression parameter and the causal effect that would be estimated in the absence of interference. 

The second problem is important in sociology as social networks are thought to reveal the structure of a group, organization, or society as a whole \cite{Free:2004}. For example, there has always been great interest in determining whether the triad is an important social unit \cite{Simm:1908,Heid:1946}. If the existence of network ties A-B and A-C makes the presence of network tie B-C more likely then the network exhibits {\it transitivity}, commonly described as ``a friend of a friend is a friend''. Thus, just as an individual may influence or be influenced by multiple others, the relationship status of one dyad (pair of individuals) may affect the relationship status of another dyad, even if no individuals are common to multiple dyads. Accounting for between dyad dependence is a core component of many social network analyses and has entailed much methodological research.

Network science is a parallel field to social network analysis in that there is very little overlap between researchers in the respective fields despite the similiarity of the problems. Whereas solutions to problems in social networks have tended to be data-oriented in that models and statistical tests are based on the data, those in network science have tended to be phenomenon-oriented with analogies to problems in the physical sciences often providing the backbone for solutions. Methods for social network analysis often have causal hypotheses (e.g., does one individual have an effect on another, does the presence of a common friend make friendship formation more likely) motivating them and involving micro-level modeling. In contrast, methods in network science seek models generated from some theoretical basis that reproduce the network at a global or system level and in so-doing reveal features of the data generating process (e.g., is the network scale-free, does the degree-distribution follow a power-law). One of the goals of this chapter is to address the lack of interaction between the social network and network science fields by providing the first joint review of both. By enlarging the range of methods at the disposal of researchers, advances at the frontier of networks and health will hopefully accelerate.

The computer age has enabled widespread implementation of methods for social network and network science analysis, particularly statistical models. At the same time, a diverse range of applications of social network analysis have appeared, including in medicine \shortcite{Keat:2007,Pham:2009,Barn:2012a,Iwas:2002,Poll:2012}. Because many medical and health-related phenomena involve interdependent actors (e.g. patients, nurses, physicians, and hospitals), there is enormous potential for social network analysis to advance health services research \cite{Omal:2008}. 

The layout of the remainder of the chapter is as follows. This introductory section concludes with a brief historical account of social networks and network science is given. The major types of networks and methods for representing networks are then discussed (Section 2). In Section 3 formal notation is introduced and descriptive measures for networks are reviewed. Social influence and social selection are studied in Sections 4 and 5, respectively. Our focus switches to methods akin with network science in Section 6, where descriptive methods are discussed. The review of network science methods continues with community detection methods in Section 7 and generative models in Section 8. The chapter concludes in Section 9.

\subsection{Historical Note}

In the 1930's, a field of study involving human interactions and relationships emerged simultaneously from sociology, psychology, and anthropology. Moreno is credited for inventing the sociogram \cite{More:1934}, a visual display of social structure. The appeal of the sociogram led to Moreno being considered a founder of sociometry, a precursor to the field known as {\it social networks}. A number of mathematical analyses of network-valued random variables in the form of sociograms followed \cite{Fest:1949,Katz:1947,Katz:1953,Katz:1955}. Other important contributions were to structural balance \cite{Heid:1946,Newc:1953,Cart:1956}; the diffusion of medical innovations \cite{Cole:1957,Cole:1966}; structural equivalence \cite{Lorr:1971}; and social influence \cite{Mars:1993}. Refer to \citeN[chapter 1]{Wass:1994} for a detailed historical account.

Early network studies involved small networks with defined boundaries such as students in a classroom, or a few large entities such as countries engaging in international trade. Because the typical number of individuals in such studies was small (e.g., $\leq 100$), relationships could be determined for all possible pairs of individuals yielding complete {\it sociocentric} datasets. Furthermore, the often enclosed nature of the system (e.g., a classroom or commune) reduced the risk of confounding by external factors (e.g., unobserved actors).

Sociological theory developed over time as sociologists provided intuitive reasoning to support various hypotheses involving social networks and society \cite{Free:2004}. In the specific area of individual health, at least five principal mediating pathways through which social relationships and thus social networks may influence outcomes have been posited \cite{Berk:2000}. Prominent among these is social support, which has emotional, instrumental, appraisal (assistance in decision making), and informational aspects \cite{Hous:1985}. Beyond social support, networks may also offer access to tangible resources such as financial assistance or transportation. They can also convey social influence by defining norms about such health-related behaviors as smoking or diet, or via social controls promoting (for example) adherence to medication regimes \cite{Mars:2006}. Networks are also channels through which certain communicable diseases, notably sexually transmitted ones, spread \cite{Klov:1985} and certain network structures have been hypothesized to reduce exposure to stressors \cite{Hain:1992}.

A field known as mathematical sociology complemented social theory by attempting to derive results using mathematical rather than intuitive arguments. In particular, statistical and probability methods are used to test for the presence of various structural features in the network. Other key areas of mathematics that have been used in network analysis include graph theory and algebraic models. \citeN{Katz:1955} develop tests of dependence within dyads (pairs of actors) while (\citeNP{Hara:1953,Hara:1955}) develop tests of triadic dependence. In general, results were descriptive or based on simple models making strong assumptions about the network. With the advent of powerful computers, mathematical contributions have taken on more importance as so much more can be implemented that in the past. For example, computer simulation has recently been used to test and develop theoretical results \cite{Cent:2009}.

In the mid-late 1990s, network science emerged as a discipline. Whereas social networks were the domain of social scientists and a growing number of statisticians, network scientists typically have backgrounds in physics, computer science or applied mathematics. The use of physical concepts to generate solutions to problems is common as evinced by the large domains of research focusing on the adaptation of (e.g.) a particular physical equation to network data. For example, several procedures for partitioning a network into disjoint groups of individuals (``communities'') rely on the {\it modularity} equation, which was developed in the context of spin-theory to model the interaction of electrons. While much of the initial work focused on the properties of the solution at different values of the parameters there recently been has increased attention to using these methods to provide valuable insight on important practical problems.

\section{Representation of networks}

Social networks are comprised of units and the relationships between them. The units are often individuals (also referred to as {\it actors}) but can include larger (e.g., countries, companies) and smaller (e.g., organisms, genes) entities.

\subsection{Network data}

In sociocentric studies, data is assembled on the ties linking all units or actors within some bounded social collective \cite{Laum:1983}. For example, the collection of data on the network of all children in a classroom or on all pairs of physician collaborations within a medical practice constitutes a sociocentric study. Relationships can be shared or directional, and quantified by binary (tie exists or not), scale (or valued), or multivariate variables. By measuring all relationships, sociocentric data constitutes the highest level of information collection and facilitates an extensive range of analyses including accounting for the effects of multiple actors on actor outcomes or the structure of the network itself to be studied \cite{Omal:2008}. A weaker form of relational data is collected in egocentric studies where individuals (``egos'') are sampled at random and information is collected on at least a sample of the individuals with direct ties to the egos (``alters''). Because standard statistical methods such as regression analysis can generally be used to analyze egocentric data \shortcite{Omal:2012}, herein egocentric data are not featured.

Relational data is often binary (e.g., friend or non-friend). One reason is that other types of relational data (e.g., nominal, ordinal, interval-valued) are often transformed to binary due to the convenience of displaying binary networks. Another is the greater range of models available for modeling binary data.

Many studies involve two distinct types of units, such as patients and physicians, or physicians and hospitals, authors and journal articles or books, etc. In these two-mode networks, the elementary relationships of interest usually refer to affiliations of units in one set with those in the other; e.g., of patients with the physician(s) responsible for their care, or of physicians with the hospital(s) at which they are admitted to practice. Two-mode networks are also known as affiliation or {\it bipartite} networks. They can be viewed as a special-case of general sociocentric network data in that the relationship of interest is between heterogeneous types of actors.

The advent of high-powered computers has enabled the analysis of large networks, which has benefitted fields such as health services research that regularly encounter large data sets. A challenge facing analyses of large networks is that it may be infeasible for all actors to be exposed to each other actor and thus for a relationship to have formed. Therefore, statistical analyses for large networks essentially use relational data representing the joint event of individuals meeting and then forming a tie, not the network of ties that would be observed if all pairs of individuals actually met. Accordingly, analyses of large networks may underestimate effect sizes unless information on the likelihood of two individuals meeting is incorporated.

\subsection{Representation of network data}

Let the status of the relationship from $i$ to $j$ be denoted by $a_{ij}$, element $ij$ of the adjacency matrix $A$. In a directed network $a_{ij}$ may differ from $a_{ji}$ while in a non-directed network $a_{ij}=a_{ji}$, implying $\bm{A}=\bm{A}^{T}$. A network constructed from friendship nominations is likely to be directed while a network of coworkers is non-directed. In the case of immutable relationships (e.g., siblings), $\bm{A}$ will only change as actors are added or removed (e.g., through birth or death), as relationship status is otherwise invariant. In the following, assume the network is binary unless otherwise stated.

Matrices and graphs are two common ways of representing the status of a networks at a fixed time. In a matrix representation, rows and columns correspond to units or actors; the matrix is square for one-mode and rectangular for two-mode networks. Elements of the matrix contain the value of the relationship linking the corresponding units or actors, so that element $ij$ represents the relationship from actor $i$ to actor $j$. With binary ties (1 = tie present, 0 = tie absent), the matrix representation is known as an adjacency matrix. Irrespective of how the network is valued, the diagonal elements of the matrix representing the network equal 0 as self-ties are not permitted. Several network properties can be computed through matrix operations.

\begin{figure}
\centering
\includegraphics[width=4.5in]{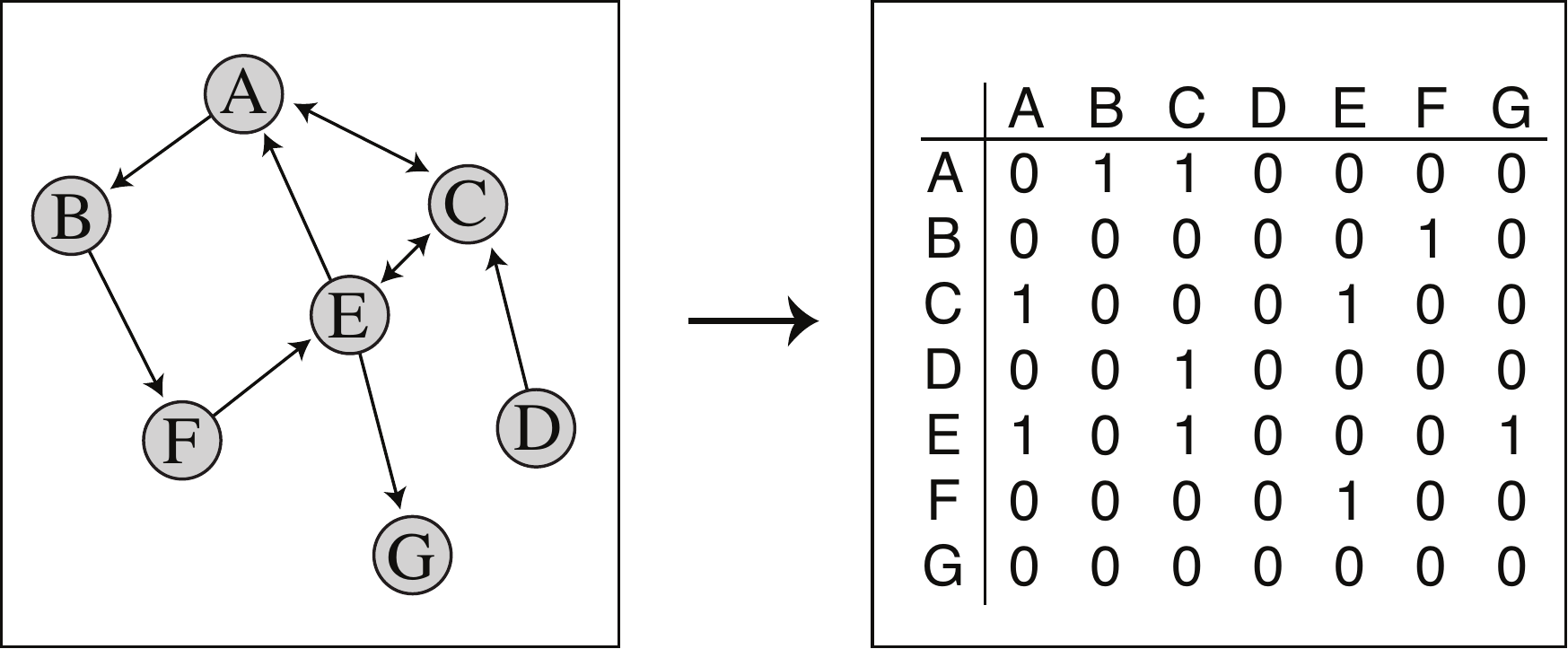}
\caption{Graphical and matrix representation of a social network. Digraph (left) and adjacency matrix (right), which is denoted in the text as $\bm{A}$. Note: Self-ties are not relevant in studies involving relationships.}
\label{fig:NetRep1}
\end{figure}

In graphical form, units or actors are vertices and non-null relationships are lines. Non-directed relationships are known as ``edges'' and directed ones as ``arcs''; arrows at the end(s) of arcs denote their directionality. Value-weighted graphs can be constructed by displaying non-null tie values along arcs or edges, or by letting thinner and thicker lines represent line values. Such graphical imagery is a hallmark of social network analysis \cite{Free:2004}.

Two-mode (or bipartite) networks may be represented in set-theoretic form as hypergraphs consisting of a set of actors of one type, together with a collection of subsets of the actors defined on the basis of a common actor of the second type \cite{Wass:1994}. This representation highlights the multi-party relationships that may exist among those actors of one type that are linked to a given actor of the other type; e.g., the set of all physicians affiliated with a particular clinic or service. In matrix form, element $ij$ of an affiliation matrix $\bm{A}$ indicates that actor $i$ of the first type is linked to actor $j$ of the second type. Affiliation networks may usefully be represented as bipartite graphs in which nodes are partitioned into two disjoint subsets and all lines link nodes in different sets.

\begin{figure}
\centering
\includegraphics[width=4.5in]{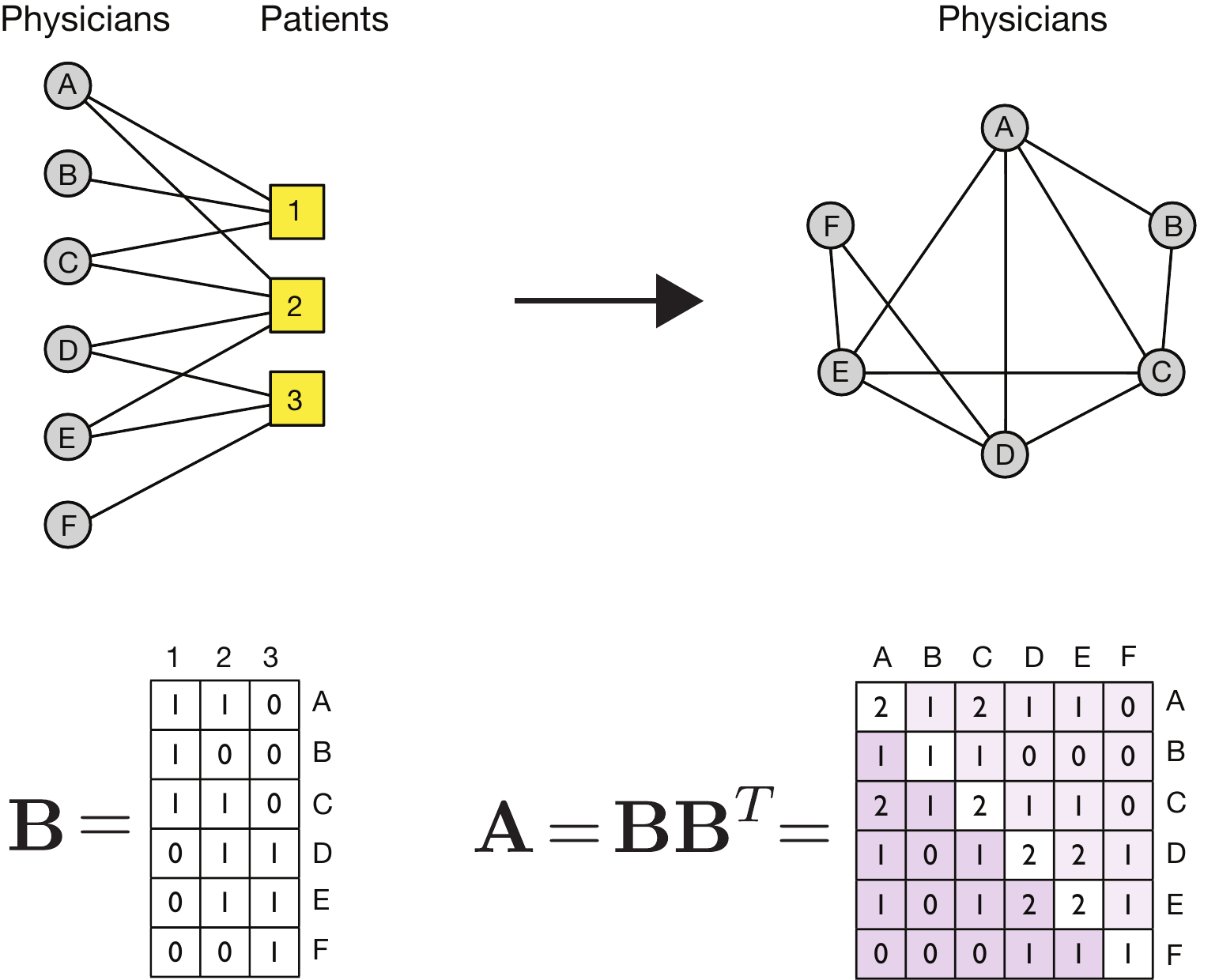}
\caption{A schematic illustrating a projection from a two-mode (bipartite) to a one-mode (unipartite) network. For example, Medicare records link each doctor to a number of patients, defining a bipartite network consisting of two types of nodes, doctors and patients. An edge can only exist between different types of nodes (a doctor and a patient), and the network is fully described by the (in this case $6 \times 3$) bipartite adjacency matrix $\bm{B}$. A one-mode projection of the doctor-patient network is obtained by multiplying the bipartite adjacency matrix $\bm{B}$ by its transpose, $\bm{B}^{T}$, to yield a $6 \times 6$ symmetric one-mode adjacency matrix $\bm{A}$, whose elements indicate the number of patients the two physicians have in common. The diagonal elements of $\bm{A}$ correspond to the number of patients the given physician "shares with themselves" (i.e., the number of patients they care for).} 
\label{fig:BiparProj}
\end{figure}

An induced one-mode network $\bm{A}$ may be obtained by multiplying an affiliation matrix $\bm{B}$ by its transpose, $\bm{A}=\bm{BB}^{T}$; entry $ij$ of the outer-product $\bm{BB}^{T}$ gives the number of affiliations shared by a pair of actors of one type (see Figure~\ref{fig:BiparProj}, which emulates a figure in \shortcite{Land:2012}). Dually, the inner-product $\bm{B}^{T}\bm{B}$ yields a one-mode network of shared affiliations among actors of the second type \cite{Brei:1974}. The diagonals of the outer and inner matrix products give the degree of the actors (i.e., the number of ties to actors of the other mode).  

In health services applications, an investigator is often interested in a one-mode network that is not directly observed but rather is induced from a two-mode network. Such one-mode projection networks are motivated theoretically by a claim that shared actors from the other mode act as surrogates for ties between the actors. For example, physicians with many patients in common might have heightened opportunities for contact through consultations or sharing of information about those patients and thus the number of shared patients is a surrogate for the actual extent of interaction between pairs of physicians. Examples of provider (physician, hospital, health service area) networks obtained as one-mode projections of bipartite networks in health services research are given in (\shortciteNP{Barn:2012a,Barn:2012b,Barn:2012c,Pham:2009}). 

An often overlooked feature of bipartite network analysis is the mechanism by which network data is obtained. Networks obtained from one-mode projections have different statistical properties from directly-observed one-mode networks. Consider a patient-physician bipartite network and suppose a threshold is applied to the physician one-mode projection such that true social ties are assumed to exist or not according to whether one or more patients are shared. Then a patient that visits three physicians is seen to induce ties between all three physicians. The same complete set of ties between the three physicians is also induced by three patients that each visit different pairs of the three physicians. However, the projection does not preserve the distinction (see Section~\ref{sec:bipar} for further comment).

\section{Descriptive Measures}

\subsection{Unipartite or one-mode networks}
\label{sec:Note}

The number of units or actors ($N$) is known as the order of the network. A common network statistic is network density ($D$), defined as the number of ties across the network ($L$) divided by the number of possible ties; for directed networks $D=L/(N(N-1))$ and for non-directed networks $D=L/(2N(N-1))$. Thus, density equals the mean value of the binary (1, 0) ties across the network. The same definition can be used for general relational data, in which case the resulting measure is sometimes referred to as {\it strength}. While results in this chapter are generally presented for binary networks, corresponding measures for weighted networks often exist (\citeNP{Opsa:2010,Opsa:2010}).

The tendency for relationships to form between people having similar attributes is known as homophily \shortcite{McPh:2001}. Homophily involves subgroup-specific network density statistics.  With high homophily according to some attribute, networks tend toward segregation by that attribute - the extreme case occurs when the network consists of separate components (i.e., no ties between actors in different components) defined by levels of the attribute. In the other direction, one obtains a bipartite network where all ties are between different types of actors (extreme heterophily).

The out- and in-degree for an actor $i$ are the number of ties from, $a_{i+}=\sum_{j=1}^{N} a_{ij}$ (column sum), and to, $a_{+j}=\sum_{i=1}^{N} a_{ij}$ (row sum), actor $i$. These are also referred to as {\it expansiveness} and {\it popularity}, respectively. For example, a positive correlation between out- and in-degree suggests that popular individuals are expansive. 

The number of ties (or value of the ties) in a network is given by $L=N \bar{d}$, where $\bar{d}$ denotes the mean degree (or strength) of an individual, implying the density of the network is given by $D=\bar{d}/(N-1)$. This result is not specific to in- or out-degree due to the fact that the total number of inward ties must equal the total number of outward ties, implying mean in-degree equals mean out-degree.

\begin{figure}
\centering
\includegraphics[width=4in]{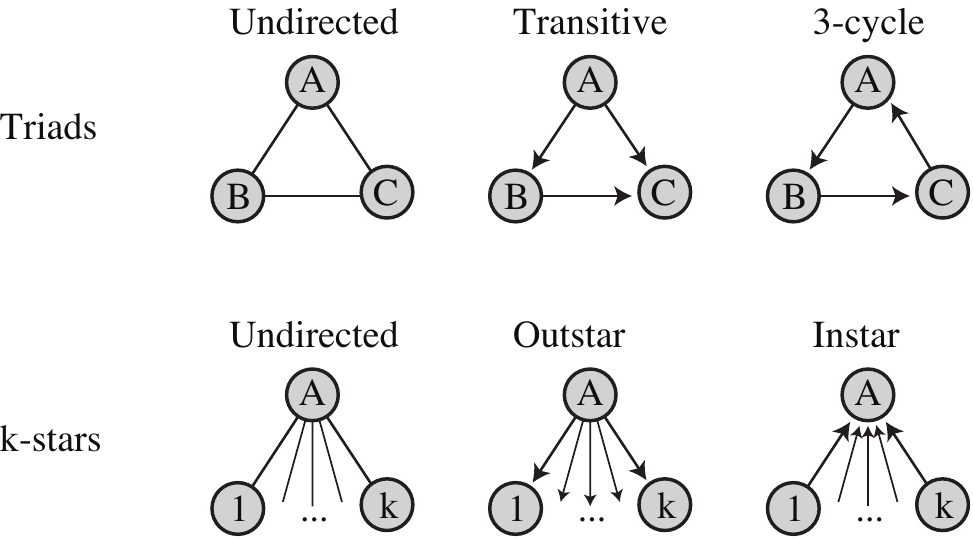}
\caption{Triadic and k-star configurations.}
\label{fig:config}
\end{figure}

The variance of the degree distribution measures the extent to which tie-density (or connectedness) varies across the network \cite{Snij:1981}. Often actors having higher degree have prominent roles in the network \cite{Free:1979}. A special type of homophily is the phenomenon where individuals form ties with individuals of similar degree, commonly referred to as {\it assortative mixing}. In directed networks, assortative mixing can be defined with respect to both out-degree and in-degree \cite{Pira:2010}. The opposite scenario to a network with the same degree for all actors is a $k$-star -- a network configuration with $k$ relationships are incident to the focal actor (Figure~\ref{fig:config}) -- in which there are no ties between the other actors.

The length of a path between two actors through the network is defined as the number of ties traversed to get from one actor to the other. The elements of the adjacency matrix multiplied by itself $k-1$ times, denoted $A^{k}$, equal the number of paths of length $k$ between any two actors with the number of $k$-cycles (including multiple or repeated loops) on the diagonal. The shortest path between two actors is referred to as the {\it geodesic distance}.

\subsubsection{Clustering}

Certain subnetworks have particular theoretical prominence. The first step-up from the trivial single actor subnetwork, also known as an isolated node, is the network comprising two actors (a ``dyad''). The presence and magnitude of a tendency toward symmetry or reciprocity in a directed network can be measured by comparing the number of mutual dyads (ties in both directions) to the number expected under a null model that does not accommodate reciprocity. If the number of mutual dyads is higher than expected, there is a tendency towards reciprocation.

A {\it triad} is formed by a group of three actors. Figure~\ref{fig:config} shows a "transitive triad", so-named as it exhibits the phenomenon that a ``friend of a friend is a friend.'' Non-parametric tests for the presence of transitivity or other forms of triadic dependence are based on the distribution of the number of closed and non-closed triads conditional on the number of null (no ties intact), directed (one tie intact), and mutual dyads (both ties intact) collectively known as the {\it dyad census}; the degree distribution; and other lower-order effects (e.g., homophily of relevant individual characteristics) in the observed network. Such tests are described in \citeN[chapter 14]{Wass:1994}. 

\subsubsection{Centrality}

Centrality is the most common metric of an actor's prominence in the network and many distinct measures exist. They are often taken as indicators of an actor's network-based ``structural power.'' Such measures are often used as explanatory variables in individual-level regression models \shortcite{Barn:2012a}. 

Different centrality measures are characterized by the aspects of an actor's position in the network that they reflect. For example, degree-based centrality -- the degree of an actor in an undirected network and in- or out-degree in a directed network -- reflects an actor's level of network connectivity or involvement in the network. Betweenness centrality computes the frequency with which an actor is found in an intermediary position along the geodesic paths linking pairs of other actors. Actors with high betweenness centrality have high capacity to broker or control relationships among other actors. A third major centrality measure, closeness centrality, is inversely-proportional to the sum of geodesic distances from a given actor to all others.  The rationale underlying closeness measures is that actors linked to others via short geodesics have comparatively little need for intermediary units, and hence have relative independence in managing their relationships.  Closeness measures are defined only for networks in which all actors are mutually related to one another by paths of finite geodesic distance; i.e., single component networks. Finally, eigenvalue centrality is sensitive to the presence or strength of connections, as well as those of the actors to which an actor is linked \cite{Bona:1987}. It assumes that connections to central actors indicate greater prominence than do (similar-strength) connections to peripheral actors. The key component of the measure is the largest eigenvalue of an adjacency or other matrix representation of the network \cite{Bona:1987}.

Network-level centrality indices \cite{Free:1979} are network-level statistics that resemble the degree variance whose values grow larger to the extent that a single actor is involved in all relationships (as in the ``star'' network shown in Figure~\ref{fig:config}). 

\subsubsection{Cliques, Components and Communities}

The assignment of actors to groups is an important and growing field within social networks. The rationale for grouping actors is that it may reveal salient social distinctions that are not directly observed. The general statistical principle adhered to is that individuals within a group are more alike than individuals in different groups. Groups are typically formed on the basis of network ties alone, the rationale being that the similarity of individuals positions in the network is in-part revealed by the pattern of ties involving them. Thus, actors in densely connected parts of the network are likely to be grouped together. A related concept to a group is a clique, a maximal subset of actors having density 1.0 (i.e., ties exist between all pairs of individuals in a binary network). The larger the clique the stronger the evidence that the collective individuals are in the same group. Grouping algorithms based on maximizing the ratio of within-group to between group ties are unlikely to split large cliques as doing so creates a lot of between group ties. However, a clique need not be its own group.

Components of a network are defined by the non-existence of any paths between the actors in them. Often a network is comprised of one large component and several small components containing few individuals. A more practical way of grouping individuals than by cliques is through $k$-connected components \cite{Whit:2001}, a maximal subset of actors mutually linked to one another by at least $k$ node-independent paths (i.e., paths that involve disjoint sets of intermediary actors who also lie within the subgraph). Such a criterion is related to $k$-coreness, a measure of the extent to which subgraphs with all internal degrees $\geq k$ occur \cite{Seid:1983} in a network.

There are several other ways for grouping the actors in a network. Model-based methods include mixed-membership stochastic block models \cite{Airo:2008} and latent-class models in which the group is treated as a categorical individual-level latent variable \cite{Hand:2007} while non-parametric methods used in network science include {\it modularity} and its variants. These methods are discussed in Section~\ref{sec:PhysCD}, where the grouping of actors is referred to as {\it community detection}.

\subsection{Bipartite or two-mode networks}
\label{sec:bipar}

In practice two-mode networks are rarely directly analyzed. If one of the modes instigates ties or is of primary interest, the network involving just those actors is often analyzed as a single-mode network. For example, in a physician-patient referral network, the physicians often instigate ties through patient referrals while patients are chiefly responsible for who they see first. The projection from a two-mode network to a one-mode network links nodes in one mode (e.g., physicians) if they share a node of the other mode (e.g., patients). A weighted network can be formed with the number of shared actors of the other mode (or function thereof) as weights.

In describing networks obtained from a projection of a two-mode network, the usual practice is to use unipartite descriptive measures. However, several layers of information are lost, including the number of actors in the other mode underlying a tie and the degree distribution of the actors in the other mode, from treating a one-mode projection as an actual network. Even if the two-mode network is completely random, ties in a one-mode projection that arise from a single (e.g.) patient with ties to (e.g.) three physicians are not separate events. More generally, a patient who visits $k$-physicians generates a $k$-clique among those physicians and tells us nothing about whether physician sharing of one patient is correlated with physician sharing of another patient -- the question of primary interest in the study of the diffusion of treatment practices. Thus, $k$-cliques for $k>2$ may be excluded from measures of transitivity in two-mode networks.

Descriptive measures for two-mode networks may be computed that parallel those for one-mode networks \cite{Wass:1994}. Centrality measures based on the bipartite network representation are covered in \citeN{Faus:1997}. \citeN{Borg:1997} review visualization, subgroup detection, and measurement of centrality for two-mode network data. More descriptive measures for two-mode networks have recently been proposed. For example, a two-mode measure of transitivity defined as the ratio of the total number of six cycles (closed paths of six ties through six nodes) in the two-mode network divided by the total number of open five-paths through six nodes \cite{Opsa:2012}. In the context of the patient-physician network, {\it physician transitivity} exists if physicians A and B sharing a patient and physicians B and C sharing a patient makes it more likely for physicians A and C to share a patient. It is only if the two pairs of physicians have different patients in common that the physician triad may be transitive and only if the third pair share a different patient from the first two that the event can be attributed to transitivity. The involvement of distinct patients makes the physician-physician ties distinct events and thus informative about clustering of physicians (and patients).

In general, the matrix equation $\bm{A}=\bm{BB}^{T}$ in which a bipartite network adjacency matrix $B$ is multiplied by its transpose yields a weighted one-mode network (the elements contain the number of shared actors of the other mode). To avoid losing information about the number of actors leading to a tie between primary nodes, weights can be retained or monotonically transformed in the projected network. Weighted analogies of descriptive measures of binary networks can be evaluated on the weighted one-mode projection. For example, the calculation of degree is emulated by summing the weights of the edges involving an individual, yielding their {\it strength}. Degree and strength together distinguish between actors with many weak ties and those with a few strong ties. Analogous measures of centrality can also be computed for the weighted one-mode projection \cite{Opsa:2010}. However, whether ties between $k$ physicians arise through them all treating the same patient, from each pair of physicians sharing a unique patient, or some in-between scenario cannot be determined post-transformation; thus, the projection transformation expends information.

A further strategy is to set weights for the bipartite network prior to forming the projection. For example, in co-authorship networks, the tie connecting an author to a publication might receive a weight of $1/(N_{j}-1)$ where $N_{j}$ is the number of authors on paper $j$ \cite{Newm:2001}. (Only papers with at least two authors are used to form such networks.) The rationale is that the greater the number of authors the lower the expected interaction between any pair (a similar logic underlies the example weight matrix described in Section~\ref{sec:Influ}). The sum of the weights across all publications common to two authors is then the basis of their relationship in the author network. 

If the events defining the bipartite network occur at different times (e.g., medical claims data often contain time-stamps for each patient-physician encounter) a directed one-mode network may be formed. The value of the A-B and B-A ties in the physician-physician network could be the number of patients who visited A before B and B before A, respectively. In the resulting directed network each physician has a flow to and from each other physician. Subsequent transformation of the flows to binary values yields dyads with states null, directed, and mutual as in a directed unipartite binary network. 

Because medical claims and surveys are frequent sources of information about one entity's experience (e.g., a patient) with another entity (e.g., a health plan or physician), bipartite network analysis is an area that promises to have enormous applicability to health services research. Hence, new methods for bipartite network analysis are needed.

\section*{Part II: Statistical Models}

We now consider the use of statistical models in social network analysis. Particular emphasis is placed on methods for estimating social influence or peer effects and models for analyzing the network itself, including accounting for social selection through the estimation of effects of homophily. 

\section{Network Influence Models}
\label{sec:Influ}

Reported claims about peer effects of health outcomes such as BMI, smoking, depression, alcohol use, and happiness have recently tantalized the social sciences. In large part, the discussion and associated controversies have arisen from the statistical methods used to estimate peer effects \cite{Omal:2013,Chri:2013}. 

Let $y_{it}$ and $\bm{x}_{it}$ denote a scalar outcome and a vector of variables, respectively, for individual $i=1,\ldots,N$ at time $t=1,\ldots,T$ ($\bm{x}_{it}$ includes 1 as its first element to accommodate an intercept). In this section, the relationship status of individuals $i$ and $j$ from the perspective of individual $i$ (denoted $a_{ij}$), is assumed to be time-invariant. For ease of notation no distinction is made between random variables and realizations of them. The vector $\bm{Y}_{t}$ and the matrices $\bm{X}_{t}$ and $\bm{A}$ are the network-wide quantities whose $i$th element, $i$th row, and $ij$th element contain the outcome for individual $i$, the vector of covariates for individual $i$, and the relationship between individuals $i$ and $j$ as perceived by individual $i$, respectively. The representation of an example adjacency matrix, denoted $\bm{A}$, is depicted in Figure~\ref{fig:NetRep1}.

Regression models for estimating peer effects are primarily concerned with how the distribution of a dependent variable (e.g. a behavior, attitude or opinion) measured on a focal actor is related to one or more explanatory variables. When behaviors, attitudes or opinions are formed in part as the result of interpersonal influence, outcomes for different individuals may be statistically dependent. The outcome for one actor will be related to those for the other actors who influence her or him, leading to a complex correlation structure. 

\begin{figure}
\centering
\includegraphics[width=4.5in]{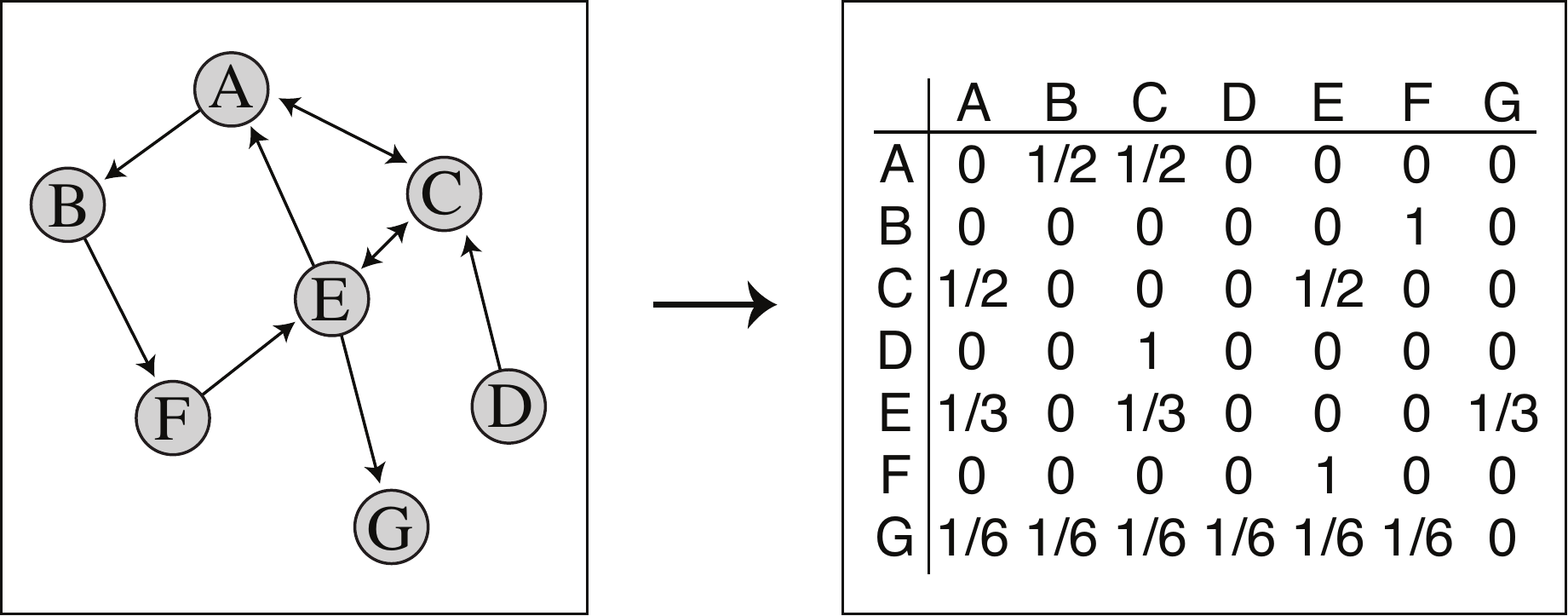}
\caption{Construction of a network weight matrix $\bm{W}$ (right). A directed edge from $i$ to $j$ means that node (or individual) $i$ has a relationship to node $j$ while element $ij$ of $\bm{W}$ quantifies the extent that individual $i$ is influenced by individual $j$. Although the mathematical form of influence depicted here assumes that influence only acts in the direction of the edge, influence may in general act in the absence of a tie (e.g., people who consider me as a friend might influence me even if I do not consider them a friend).}
\label{fig:NetRep2}
\end{figure}

In social influence analyses the weight matrix, $\bm{W}=[w_{ij}]$ in Figure~\ref{fig:NetRep2}, apportions the total influence acting on an individual evenly across the individuals with whom they have a netwok tie. Typically
\begin{enumerate}
\item $w_{ij} \geq 0$: non-negative weights.
\item $w_{ii}=0$: no self-influence.
\item $\sum_{j} w_{ij}=1$: weights give relative influences (because its row-sums equal 1, $\bm{W}$ is said to be row-stochastic).
\end{enumerate}
Let $\bar{y}_{-it}=(\bm{W}\bm{Y}_{t})_{i}$ denote the influence-weighted average of the outcome $y$ across the network after excluding (i.e., subtracting) individual $i$ from the set of individuals to be averaged over. Similarly, let $\bm{\bar{x}}_{-it}^{T}=(\bm{W}\bm{X}_{t})_{i}$ denote the vector containing the corresponding influence weighted covariates, often referred to as {\it contextual variables}.

The most common choice for $\bm{W}$ is the row-stochastic version of $\bm{A}$. For illustration, suppose that $\bm{A}$ is binary (the elements are 1 and 0). Then the off-diagonal elements on the $i$th row of $\bm{W}$ equal $a_{i+}^{-1}$ if $a_{i+}>0$ and $1/(N-1)$ otherwise (Figure~\ref{fig:NetRep2}). This framework assumes that an individual's alters are equally influential. In general, influence might only transmit through outgoing ties (e.g., those individuals viewed as friends by the focal actor - a scenario consistent with Figure~\ref{fig:NetRep2}), or might only transmit through received ties (e.g., individuals who view the focal actor as a friend), or might act in equal or different magnitude in both directions.

Network-related interdependence among the outcomes may be incorporated in two distinct ways. First, an outcome for one actor may depend directly on the lagged outcomes or lagged covariates of the alters to whom she or he is linked. For example, consider the model:
\begin{equation}
 y_{it} = \al_{1}\bar{y}_{-i(t-1)} + \bm{\al}_{x}^{T}\bm{\bar{x}}_{-i(t-1)} + \be_{1}y_{i(t-1)} + \bm{\be}_{2}^{T}\bm{x}_{i(t-1)} + \ep_{it}, \label{eq:Dynam1}
\end{equation}
where $\al_{1}$ is a scalar parameter quantifying the peer effect; $\bm{\al}_{x}$ is a $p$-dimensional vector of parameters of peer effects acting through the $p$ covariates in $\bm{x}$, $\bm{\be}=(\be_{1},\bm{\be}_{2}^{T})^{T}$ is a vector of other regression parameters for the within-individual predictors, and $\ep_{it}$ is the independent error assumed to have mean 0 and variance $\var$. The notation used in Equation~\ref{eq:Dynam1} is adopted through this section; hence, $\al$ and $\be$ denote peer effects and within-individual effects, respectively. 

Equation (\ref{eq:Dynam1}) is known as the ``linear-in-means model'' \cite{Mans:1993} due to the conduit for peer influence being the trait averaged over the alters of each focal actor. The model has a symmetric appearance in that it contains corresponding peer effects for each of the within individual predictors. A common alternative model assumes $\bm{\al}_{x}=0$; in other words, that peer effects only act through the same variable in the alters as the outcome. Another set of variants arises in the case when there are multiple types of alters with heterogeneous peer effects. Such a situation may be represented in a model by defining distinct influence matrices for each type of peer. Let $\bm{W}^{(h)}$ denote the weight matrix formed from the adjacency matrix for the network comprising only alters of type $h$ and let $\bar{y}_{-i(t-1)}^{(h)}=(\bm{W}^{(h)}\bm{Y}_{t-1})_{i}$ for $h=\{1,\ldots,H\}$, where $H$ is the number of distinct types of alters. Then an extension of the linear-in-means model to accommodate heterogeneous peer effects is:
\begin{equation}
 y_{it} = \sum_{h=1}^{(h)} \delim{(}{ \al_{1}^{(h)}\bar{y}_{-i(t-1)}^{(h)} + (\bm{\al}_{x}^{(h)})^{T}\bm{\bar{x}}_{-i(t-1)} }{)} + \be_{1}y_{i(t-1)} + \bm{\be}_{2}^{T}\bm{x}_{i(t-1)} + \ep_{it}. \label{eq:Dynam2}
\end{equation}
In the special case where $\al_{1}^{1}=\al_{1}^{2}=\ldots=\al_{1}^{(h)}$ and $\bm{\al}_{x}^{1}=\bm{\al}_{x}^{2}=\ldots=\bm{\al}_{x}^{(h)}$, (\ref{eq:Dynam2}) reduces to (\ref{eq:Dynam1}). An alternative to (\ref{eq:Dynam2}) is to fit separate models for each type of peer, which would yield estimates of the overall (or marginal) peer effect for each type of peer as opposed to the independent effect of each type of peer above and beyond that of the other types.

\begin{figure}
\centering
\includegraphics[width=2in]{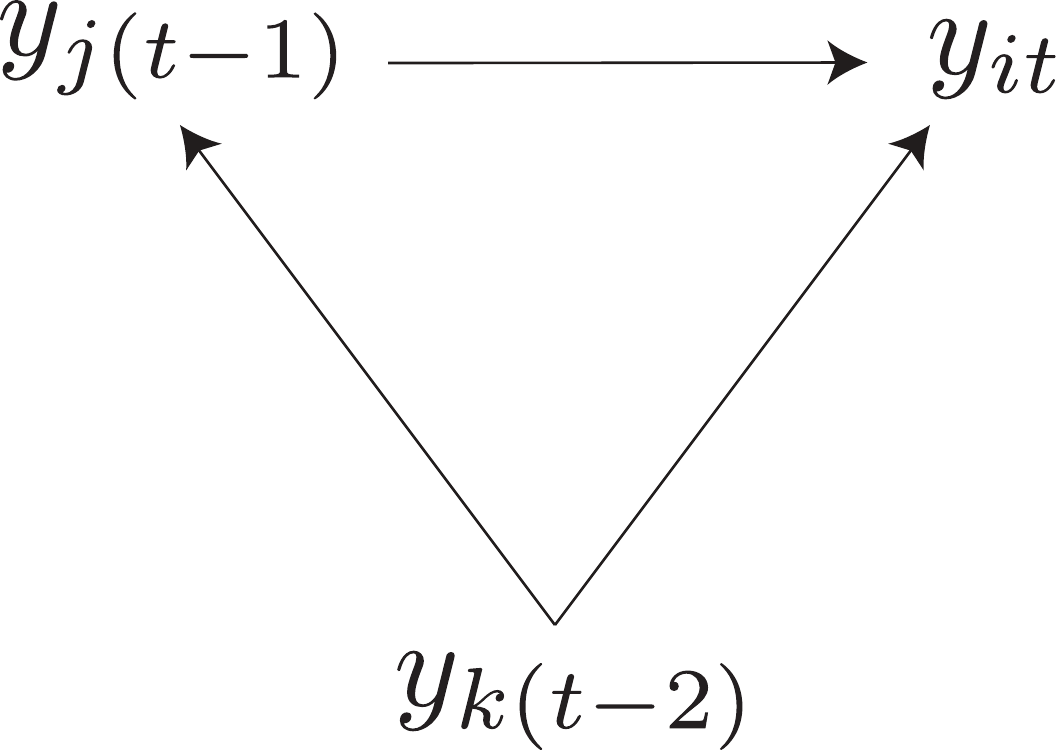}
\caption{Simplified directed acyclic graph (DAG) illustrating confounding of a peer effect by a third individual. The DAG is simplified because it does not explicitly show the variable $y_{k(t-1)}$, which is an intermediary between $y_{k(t-2)}$ and $y_{it}$. (Because the point made here does nut depend on $y_{j(t-2)}$ and $y_{i(t-1)}$ they are not depicted.) If $y_{k(t-2)}$ (or $y_{k(t-1)}$) is conditioned on, the path $y_{j(t-1)} \leftarrow y_{k(t-2)} (\rightarrow y_{k(t-1)}) \rightarrow y_{it}$ is unblocked and therefore confounds $y_{j(t-1)} \rightarrow y_{it}$, whose effect is the peer effect of interest. Although the DAG looks like a digraph of a network, a DAG is a different construction.} 
\label{fig:NetConfound}
\end{figure}

Failing to account for all alters may lead to biased results if the alters are interconnected. Figure~\ref{fig:NetConfound} presents a simple directed acyclic graph (DAG), which is a device for determining whether or not an effect is identifiable, involving three individuals $i$, $j$ and $k$. The nodes represent the variables of interest (a trait measured on each individual such as their BMI) and the arrows represent causal effects (the origin of the arrow is the cause and the tip is the effect). Consider the peer effect of individual $j$ at $t-1$ on individual $i$ at $t$. A causal effect is identifiable if it is the only unblocked path between two variables. Because individual $k$ is a cause of both individual $j$ and individual $i$, the peer effect of $j$ on $i$ will be confounded by individual $k$ unless unless the analysis conditions on $y_{k(t-2)}$. 

The scenario depicted in Figure~\ref{fig:NetConfound} does not present any major difficulties as long as effects involving individual $k$ are accounted. However, if individual $k$ is not known about or is ignored, then the analysis may be exposed to unmeasured confounding. This point has particular relevance to social network analyses as networks are often defined by specifying boundaries or rules for including individuals as opposed to being finite, closed systems \cite{Laum:1983}. In situations where such boundaries break true ties, influential peers may be excluded, potentially leading to biased results.

\subsection{Estimation of Contemporaneous Peer Effects}

From a practical standpoint, it may be infeasible to use a model with only lagged predictors such as (\ref{eq:Dynam1}). For instance, the time points might be so far apart that statistical power is severely compromised. Therefore, it is tempting to use a model with contemporaneous predictors such as:
\begin{equation}
 y_{it} = \al_{0}\bar{y}_{-it} + \al_{1}\bar{y}_{-i(t-1)} + \bm{\al}_{x}^{T}\bm{x}_{-it} + \be_{1}y_{i(t-1)} + \bm{\be}_{2}^{T}\bm{x}_{it} + \ep_{it}, \label{eq:DynamCF}
\end{equation}
where adjusting for $\bar{y}_{-i(t-1)}$ seeks to isolate the peer effect acting since $t-1$. However, because $\bar{y}_{-it}$ is correlated with the outcome variables of other observations, OLS will be inconsistent. Therefore, methods to account for the endogeneity arising from the correlation between $\bar{y}_{-it}$ and $\ep_{jt}$ for $j \neq i$ -- in network science parlance the state of $\bar{y}_{-it}$ is said to be an internal product or consequence of the system as opposed to an external (exogenous) force. 

In \citeN{Chri:2007}, the most widely cited of the Christakis-Fowler peer effect papers, the endogeneity problem is resolved using a novel theoretical argument. They purported that it is reasonable to assume in a friendship network that the influence acting on the focal actor (the ego) is greatest for mutual friendships, followed by ego-nominated friendships, followed by alter-nominated friendships, and finally is close to 0 in dyads with no friendships. Furthermore, they reasoned that because unmeasured common causes should affect each dyad equally. Because the estimated peer effects declined from large and positive for mutual friendships to close to 0 for alter and null friendships, consistent with their theory, it was suggested that this constituted strong evidence of a peer effect. Despite the compelling argument, \citeN{Shal:2011} revealed that unobserved factors affecting tie-formation (homophily) may confound the relationship and thus lead to biased effects. The estimation of peer effects is a topic of ongoing vigorous debate in the academic and the popular press. Alternative approaches to the theory-based approach of Christakis and Fowler are now described.

A parametric model-based solution to endogenous feedback is to specify a joint distribution for $\bm{\ep}_{t}=(\ep_{1t},\ldots,\ep_{Nt})$. Then the reduced form of the model satisfies $\bm{Y}_{t}=\al_{0} \bm{WY}_{t} + \al_{1} \bm{WY}_{t-1} + \bm{WX}_{t-1}\bm{\al}_{x} + \be_{1}\bm{Y}_{t-1} + \bm{X}_{t-1}\bm{\be}_{2} + \bm{\ep}_{t}$ for $\bm{Y}_{t}$ to yield $\bm{Y}_{t}= (\bm{I} - \al_{0} \bm{W})^{-1}\{\al_{1} \bm{WY}_{t-1} + \bm{WX}_{t-1}\bm{\al}_{x} + \be_{1}\bm{Y}_{t-1} + \bm{X}_{t-1}\bm{\be}_{2} + \bm{\ep}_{t} \}$. The resulting model emulates a spatial autocorrelation model \cite{Anse:1988}. One way of facilitating estimation is by specifying a probability distribution for $\bm{\ep}_{t}$. However, relying on the correctness of the assumed distribution for identification may make the estimation procedure sensitive to an erroneous assumed distribution. 

A semi-parametric solution is to find an instrumental variable (IV), $z_{it}$; a variable that is related to $\bar{y}_{-it}$ but conditional on $\bar{y}_{-it}$ and $(\bar{y}_{-i(t-1)},\bm{x}_{-it}, y_{i(t-1)}, \bm{x}_{it})$ does not cause $y_{it}$. If $\bar{x}_{-it}$ is excluded from (\ref{eq:DynamCF}), its elements can potentially be used as IVs \cite{Flet:2008}. However, IV methods can be problematic if the instrument is weak or if the assumption that the IV does not directly impact $y_{it}$ (the exclusion restriction) is violated, an untestable assumption. Thus, in fitting a model with contemporaneous peer effects, one faces a choice between assuming a multivariate distribution holds, relying on the non-existence of unmeasured confounding variables, or relying on the validity of an IV. None of these assumptions can be evaluated unconditionally on the observed data.

While joint modeling and IV methods provide theoretical solutions to the estimation of contemporaneous peer effects, the notion of causality is philosophically challenged when the cause is not known to occur prior to effect. Therefore, longitudinal data provide an important basis for the identification of causal effects, in particular in negating concerns of reverse causality. If the observation times are far apart the use of lagged alter predictors may, however, substantially reduced the power of an analysis.

\subsection{Dyadic Influence Analyses}
\label{sec:DyadInflu}

If the dyads consist of mutually exclusive or isolated pairs of actors there are no inter-dyad ties and influence only acts within dyads. An example of such a situation occurs when individuals can have exactly one relationship and the relationship is reciprocated, as is the case with spousal dyads. The network influence models of Section~\ref{sec:Influ} reduce to {\it dyadic influence models} in which the predictors are based on individual alters. For example, the dyadic influence model analogous to (\ref{eq:DynamCF}) is obtained by replacing the subscript $-i$ with $j$. That is, 
\begin{equation}
 y_{it} = \al_{0}\bar{y}_{jt} + \al_{1}\bar{y}_{j(t-1)} + \bm{\al}_{x}^{T}\bm{x}_{jt} + \be_{1}y_{i(t-1)} + \bm{x}_{it}^{T}\bm{\be}_{2} + \ep_{it}. \label{eq:Dyad1}
\end{equation}
The model in (\ref{eq:Dyad1}) may be estimated using generalized estimating equations (GEE), avoiding specifying a distribution for $\ep_{it}$. However, if any relationships are bidirectional, standard software packages will yield inconsistent estimates of the peer effects as they do not account for the statistical dependence introduced by individuals who play the dual role of ego and alter at time $t$ \cite{Vand:2012}.

\subsection{Frontiers in Social Influence}
\label{sec:front}

There has recently been a lot of interest and discussion concerning causal peer effects. Issues that have been discussed include the use of ordinary least squares (OLS) for the estimation of contemporaneous peer effects \cite{Lyon:2011} and the identification of peer effects independent of homophily \cite{Shal:2011}. The discussion has helped elevate social network methodology to the forefront of many disciplines. For example, \citeN{Vand:2012} show that OLS still provides a valid test of the null hypothesis that the peer effect is zero when the true peer effect is zero. Therefore, OLS can be used to test for peer effects despite the fact that OLS estimates are inconsistent under the alternative hypothesis.

\citeN{Chri:2007} use tie directionality to account for unmeasured confounding variables under the assumption that their effect on relationship status is the same for all types of relationships. The rationale is that the estimated peer effect in dyads where the relationship is not expected to be conducive to peer influence (``control relationships'') provides a baseline against which to identify the peer effect for other types of relationships. However, this test fails to offer complete protection against unmeasured homophily \cite{Shal:2011}, reflecting the vulnerability of observational data to unmeasured sources of bias. However, sensitivity analyses that evaluate the effect-size needed to overturn the results may be conducted to help support a conclusion by illustrating that the confounding effect must be implausibly large to reverse the finding \cite{Vand:2011}.

Instrumental variable (IV) methods have also been used to estimate peer effects. A common source of instruments is alters' attributes other than the one for which the peer effect is estimated \cite{Flet:2008,Flet:2009}. Potential IVs must predict the attribute of interest in the alter but must not be a cause of the same attribute in other individuals. Attributes that are invisible such as an individuals genes appear to be ideal candidate genes. For instance, an individual with two risk alleles of an obesity gene is at more risk of increased BMI but conditional on that individual's BMI their obesity genes should not affect the BMI of other individuals. However, if the obesity genes are revealed through another behavior (a phenomenon known as {\it pleiotropy}) that is associated with BMI then, unless such factors are conditioned on, genes will not be valid IVs.

\section{Relational Analyses}
\label{sec:Relat}

Sociocentric network studies assemble data on the ties representing the relationship linking a set of individuals, such as all physicians within a medical practice. Models for such data posit that global network properties are the result of phenomena involving subgroups of (most commonly) four or fewer actors \cite{Robi:2005}. Examples of such regularities are actor-level tendencies to produce or attract ties (homophily and heterophily), dyadic tendencies toward reciprocity, and triadic tendencies toward closure or transitivity. A relational model, in essence, specifies a set of micro-level rules governing the local structure of a network. In this section, models for cross-sectional relational data are consider first followed by longitudinal counterparts of them.

The simplest models for sociocentric data assume dyadic independence. Under the random model, all ties have equal probability of occurring and the status of one has no impact on the status of another \cite{Erdo:1959}. More general dyadic models were developed in \citeN{Holl:1981} and later were extended in \citeN{Wang:1987}. Because independence is still assumed between dyads, the information from the data about the model parameters accumulates in the form of a product of the probability densities for the status of the dyadin observation over each dyad:
\begin{equation}
 L = \prod_{i < j}^{N} {\rm pr}(a_{ij},a_{ji} \mid \bm{\al}, \bm{\ga}, \bm{x}_{ij}, \bm{x}_{ji}), \label{eq:LikeFn}
\end{equation}
where $\bm{\al}=(\al_{1},\ldots,\al_{N})^{T}$ and $\bm{\ga}=(\ga_{1},\ldots,\ga_{N})^{T}$ are vectors of actor-specific parameters representing the actors' expansiveness (propensity to send ties) and popularity (propensity to receive ties), respectively, and $\bm{x}_{ij}$ is a vector of covariates relevant to $a_{ij}$ (this may include covariates specific to either actor and combined traits of both actors). It is important to realize that covariates can be directional; thus, $\bm{x}_{ij}$ need not equal $\bm{x}_{ji}$. Although the model may include other parameters, $\bm{\al}$ and $\bm{\ga}$ play an important role in network analysis due to their relationship to the degree distribution of the network and so are explicitly denoted.

When relationship status is binary, the distribution of $(a_{ij},a_{ji})$ is a four-component multinomial distribution. The probabilities are typically represented in the form of a generalized logistic regression model (an extension of the logistic regression model to $\geq 2$ categories) having the form
\begin{equation}
 {\rm pr}(a_{ij},a_{ji} \mid \bm{\al}, \bm{\ga}) = k_{ij}^{-1}\exp(\mu_{ij}a_{ij} + \mu_{ji}a_{ji} + \rho_{ij}a_{ij}a_{ji} ), \label{eq:Dindep}
\end{equation}
where
\[ {\kappa}_{ij} = 1 + \exp(\mu_{ij}) + \exp(\mu_{ji}) + \exp(\mu_{ij} + \mu_{ji} + \rho_{ij}), \]
and $\mu_{ij}$, $\mu_{ji}$ and $\rho_{ij}$ are functions of $(\al_{i},\al_{j},\ga_{i},\ga_{j})$ and $(\bm{x}_{ij},\bm{x}_{ji})$. The term $\mu_{ij}$ includes factors associated with the likelihood that $a_{ij}=1$ but not necessarily the likelihood that $a_{ji}=1$. In an non-directed network the predictors can be directional and so it is likely that $\mu_{ij} \neq \mu_{ji}$. However, the only covariates included in $\rho_{ij}$ must be non-directional as they affect the likelihood of $(a_{ij},a_{ji})=(1,1)$; the sign of $\rho_{ij}$ indicates whether a mutual tie is more (if $\rho_{ij}>0$) or less (if $\rho_{ij}<0$) likely to occur than predicted by the density terms and so is a measure of {\it reciprocity} or {\it mutuality}. Null mutuality is implied by $\rho_{ij}=0$.

In dyadic models, the terms $\mu_{ij}$, $\mu_{ji}$ and $\rho_{ij}$ account for the local network about actors $i$ and $j$ through the inclusion of $(\al_{i},\al_{j},\ga_{i},\ga_{j})$. Furthermore, other effects can be homogeneous across actors or actor-specific. For example, the p$_{1}$ model \cite{Holl:1981} assumes $\mu_{ij}=\mu+\al_{i}+\ga_{j}$ and $\rho_{ij}=\rho$, implying the covariate-free joint probability density function of the network given by
\[ p_{1}(\bm{A}) \propto \exp\delim{\{}{ \mu s_{1}(\bm{A}) + \sum_{i}^{N} \al_{i} s_{2i}(\bm{A}) + \sum_{j}^{N} s_{3j}(\bm{A}) + \rho s_{4}(\bm{A}) }{\}}, \]
where $s_{1}(\bm{A})=\sum_{i \neq j}a_{ij}$, $s_{2i}(\bm{A})=a_{i+}$, $s_{3j}(\bm{A})=a_{+j}$, and $s_{4}(\bm{A})=\sum_{i \neq j}a_{ij}a_{ji}$. Thus, the p$_{1}$ model depends on $2N+2$ network statistics and associated parameters. If the p$_{1}$ model holds within (ego, alter)-shared values of categorical attributes, a stochastic block model is obtained by allowing block-specific modifications to the density and reciprocity of ties \cite{Fine:1981,Holl:1983,Wang:1987}. An extension would allow reciprocity to also vary between blocks. Because the stochastic blockmodel extension of the p$_{1}$ model is saturated at the actor-level due to the expansiveness and popularity fixed effects, no assumption is made about differences in the degree-distributions of the actors in different blocks. Stochastic block models are the basis of mixed-membership and other recent statistical approaches for node-partitioning social network data \cite{Gold:2009,Choi:2010,newman2011}. Individuals in the same block of a stochastic block model are often referred to as being structurally equivalent.

\subsection{Models of Networks as Single Observations}
\label{sec:ergm}

A criticism of dyadic independence models is that they fail to account for interdependencies between dyads. The $p^{*}$ or exponential random graph model (ERGM) generalizes dyadic independence models \cite{Fran:1986,Wass:1996}. An ERGM has the form
\begin{equation}
{\rm Pr}(\bm{A}; \bm{\tht}) = \kappa(\bm{\tht})^{-1}\exp(\sum_{k} \tht_{k} s_{k}(\bm{A})), \label{eq:ergm}
\end{equation}
where $\bm{A}$ denotes a possible state of the network, $s_{k}(\bm{A})$ denotes a network statistic evaluated over $\bm{A}$ (e.g., the number of ties, the number of reciprocated ties), $\kappa(\bm{\tht})=\sum_{\bm{A} \in {\bm{\cal A}}}\exp(\sum_{k} \tht_{k} s_{k}(\bm{A}))$ and ${\bm{\cal A}}$ is the set of all $2^{N(N-1)}$ possible realizations of a directed network. In general, the scale factor $\kappa(\bm{\tht})$ that sums over each distinct network does not factor into a product of analogous terms. As a result, it is computationally infeasible to exactly evaluate the likelihood function of dyadic dependent ERGMs for even moderately-sized $N$ (e.g., $N>20$ is problematic \cite{Hunt:2006}). The key feature of the p$_{1}$ model that allows the probability of the network to decompose into the product of dyadic-state probabilities is that it only depends on network statistics $s_{k}(\bm{A})$ that sum individual ties or pairs of ties from the same dyad. 

If dyads are independent unless they share an actor, the network is a {\it Markov Random Graph} \cite{Fran:1986}. Markov Random Graphs may include terms for density, reciprocity, transitivity and other triadic structures, and $k$-stars (equivalent to the degree distribution) -- these terms contain sums of the products of no more than three ties. Such terms may be multiplied with actor attribute variables to define interaction effects. (An interaction is the effect of the product of two or more variables; e.g., if males and females have different tendencies to reciprocate ties then gender is said to interact with reciprocity.)

Networks that extend Markov Random Graphs by allowing four-cycles but no fifth- or higher-order terms are {\it partially conditionally dependent}. In such networks, a sufficient condition for dependence of $a_{ij}$ and $a_{kl}$ is that $a_{ik}=a_{jl}=1$ or $a_{il}=a_{jk}=1$ \shortcite{Wang:2009}. Thus, two edges may be dependent despite not having any actors in common. Partial conditional dependence is the basis of the new parameterizations of network statistics developed by \cite{Snij:2006} that have led to better fitting ERGMs (see below). 

Under ERGMs, the conditional likelihood of each tie given the other ties in the network has the logistic form:
\begin{equation}
 {\rm Pr}(a_{ij}=1 \mid \bm{A}_{ij}^{c}) = [1 + \exp(\bm{\tht}^{T}\de(\bm{A}_{ij}^{c}))]^{-1}, \label{eq:condlike}
\end{equation}
where $\bm{A}_{ij}^{c}$ is $\bm{A}$ with $a_{ij}$ excluded, $\de(\bm{A}_{ij}^{c})=S(\bm{A}_{ij}^{+})-S(\bm{A}_{ij}^{-})$ is the vector of changes in network statistics that occur if $a_{ij}$ is 1 rather than 0. Thus, the parameters of an ERGM are interpreted as the change in the log of the odds that the tie is present to not being present conditional on the status of the rest of the network \cite{Snij:2006}. A large positive parameter suggests that more configurations of the type represented in the network statistic appear in the observed network more often than expected by chance, all else equal \cite{Robi:2009}.

Due to the factorization of the likelihood function in (\ref{eq:LikeFn}), likelihood-based estimators of dyadic independence models have desirable statistical properties such as consistency and statistical efficiency. However, if the model for the network includes predictors based on three or more actors, no such factorization occurs and Markov chain Monte Carlo (MCMC) is required to optimize the likelihood function for (\ref{eq:ergm}), which for each observation involves making computations on $k^{N(N-1)/2}$ ($k=4$ if directed and $k=2$ if non-directed) distinct networks. ERGMs have been demonstrated to be estimable on networks with $N \approx 1600$ \cite{Good:2007}, but computational feasibility depends on the terms in the model and the amount of memory available. The ergm (``Exponential Random Graph Model'') package that is part of the Statnet suite in R, developed by the Statnet project, estimates ERGMs \shortcite{Hand:2010}. 

Other estimation difficulties include failure of the optimization algorithm to converge and the fitted model producing nonsensical ``degenerate'' predicted networks. {\it Degeneracy} arises because for certain specifications of $s_{k}(\bm{A})$ the network statistics are highly collinear or there is unaccounted effect heterogeneity across the network. As a result, under the fitted model the local neighborhood of networks around the observed network may have probability close to 0 and those networks with positive probability (often the empty and complete graphs) may be radically different from each other and thus the observed network \shortcite{Hand:2003,Robi:2007}. Although the average network over repeated draws has similar network statistics to the observed network, the individual networks generated under the fitted model do not bear any resemblence to the observed network.

Because an actor of degree $m$ contributes $k$-stars for $k \leq m$, $k$-star configurations are nested within one another and thus are highly correlated. Therefore, when multiple $k$-stars are predictors, extensive collinearity results. However, the estimated coefficients of successive $k$-star configurations (e.g., 2-star, 3-star, 4-star) tend to decrease in magnitude and have alternating signs, an observation often seen when multiple highly colinear variables are included in a regression model. This observation led to the development of the {\it alternating $k$-star} \cite{Snij:2006}, given by
\[ {\rm AS}(\la) = \sum_{k=2}^{N-1} (-1)^{k} \frac{S_{k}}{\la^{k-2}} \mbox{ for $\la > 1$}, \]
where $S_{k}$ denotes the number of $k$-stars, being used in place of multiple individual $k$-star terms in (\ref{eq:ergm}). A positive estimate of the coefficient of AS($\la$) suggests that the degree distribution is skewed towards higher degree nodes while a negative coefficient implies large degrees are unlikely. The value of $\la$ can be specified or estimated from the data \cite{Hunt:2007}.

Network statistics for triadic configurations -- the triangle (a non-directed closed triad) in non-directed networks and transitive triads, three-cycles, closed three-out stars, closed three-in stars in directed networks -- are the most prone to degeneracy. One reason is that heterogeneity in the prevalence of triads across the network, leads to heterogeneity in the density of ties across the network \cite{Robi:2009}. A model that assumes homogeneous triadic effects across the network is unable to describe networks with regions of high and low density; the generated networks are either dominated by excessive low density regions or by excessive high density regions. This observation suggests a hierarchical modeling strategy where the first step is to use a community detection algorithm (see Section~\ref{sec:PhysCD}) to partition the network into blocks of nodes. Then fit an ERGM (or other model) to the sub-network corresponding to each community, allowing the network statistics to have different effects within each community. The just-described modeling strategy combines methods of network science and social network analysis.

A similar approach has been used to overcome severe computational difficulties that often occur when one or multiple triadic (triangle-type) terms are included in the model. A $k$-triangle is a set of $k$ triangles resting on a common base. For example, if individuals $i$, $j$, and $k$ are one closed triad and individuals $i$, $j$, and $l$ are another then the four individuals form a 2-triangle with the edge $y_{ij}$ common to both. Let $T_{k}$ denote the number of $k$-triangles in the network. Thus, $T_{1}$ denotes the total number of closed triads, $T_{2}$ the total number of 2-triangles, and so on. The {\it alternating $k$-triangle statistic}
\[ {\rm AT}(\la) = 3T_{1} + \sum_{k=1}^{N-3} (-1)^{k} \frac{T_{k+1}}{\la^{k}} \mbox{ for $\la > 1$}, \]
was developed to perform for triadic structures what AS($\la$) performs for $k$-stars \cite{Snij:2006}. The presence of $\la$ makes AT($\la$) nonlinear in the triangle count, giving lower probability to highly clustered structures. By making the number of actors who share $k$ partners the core term, AT($\la$) can be re-written as a geometrically weighted edgewise shared partner (GWESP) statistic \cite{Good:2007,Hunt:2007}.

The AS($\la$) and AT($\la$) statistics do not differentiate between outward and inward ties. Recently, directed forms of these statistics have been introduced \cite{Robi:2009}. The directed versions of the $k$-star are threefold, corresponding to two-paths, shared destination node (activity), shared originator node (popularity). The directed versions of the $k$-triangle represent transitivity, activity closure, popularity closure, and cyclic-closure. 

\subsubsection{Bipartite ERGMs}

An alternative approach to modeling a one-mode projection (by construction a non-directed network) from a two-mode network is to directly model the two-mode network. An advantage of direct modeling is that all the information in the data is used. ERGMs or any other model applied to bipartite data need to account for the fact that ties can only form in dyads including one actor from each mode. In a dyadic independence model this is recognized simply by excluding all same mode dyads from the dataset. In general, the denominator $\kappa(\bm{\tht})$ in (\ref{eq:ergm}) only sums over networks in which there are no within mode ties. If the number of actors in the two modes are $N$ and $M$, there are $2^{NM}$ distinct non-directed networks.

The density and degree distributions may be represented in a bipartite ERGM as in a unipartite ERGM. However, with two modes it may be that two types of each network statistic and other predictor is needed. Representations of homophily in two-mode networks are defined across modes. Likewise, because there are no within mode ties, statistics that account for closure must also depend only on inter-mode ties. 

The smallest closed structure in a bipartite graph is a four-cycle (closed four-path). An example of a four-cycle is the path A--1--C--2--A in Figure~\ref{fig:BiparProj}; it includes four distinct actors and four edges are traversed to return to the initial actor. A simple measure of closure contrasts the number of closed four-cycles out of all three paths containing four unique actors with the overall density of ties. A simple model for testing whether clustering (closure) is present in a bipartite network includes density, both sets of $k$-stars, three-path, and four-cycle statistics as predictors. A significant positive effect of the four-cycle statistic suggests that two actors of degree two in one mode that have one of the actors in the other mode in common are more likely to also have the second actor in common, relative to two randomly selected actors of degree two from the same mode. For example, in a physician-patient network, clustering implies having one patient in common increases the likelihood of having another patient in common. Physicians A and C both have patients 1 and 2 in common, hence they provide evidence for bipartite closure. However, physicians E and F have patient 3 in common; despite being eligible to exhibit bipartite closure they do not, hence they provide evidence against bipartite closure.

Analogies of ERGMs and solutions to problematic issues exist for bipartite networks. For example, to avoid problems of high colinearity between the $k$-star terms, alternating $k$-star statistics can be used in place of them \shortcite{Wang:2009}. Let $S_{D}(\bm{B})$ denote the number of ties from one mode to the other, $AS_{1}(\bm{B})$ and $AS_{2}(\bm{B})$ denote the alternating $k$-star statistics for each mode, $S_{3P}(B)$ denotes the number of three-paths, and $S_{4C}(\bm{B})$ denote the number of closed four-cycles for a network $\bm{B}$. The resulting bipartite ERGM for $\bm{B}$ has the form:
\begin{equation}
{\rm Pr}(\bm{B}; \bm{\tht}) = \kappa(\bm{\tht})^{-1}\exp(\tht_{0}S_{D}(\bm{B}) + \tht_{1}AS_{1}(\bm{B}) + \tht_{2}AS_{2}(\bm{B}) + \tht_{3}S_{3P}(\bm{B}) + \tht_{4}S_{4C}(\bm{B})), \label{eq:bergm}
\end{equation}
where $\kappa(\bm{\tht})$ sums over the $MN$ possible bipartite graphs. The statistic $S_{4C}(\bm{B})/S_{3P}(\bm{B})$ is the proportion of times that two patients each visit the same two physicians out of all the occurrences where two patients both have one visit to one physician and one patient visits the other physician. The coefficient $\tht_{4}$ is the effect associated with this lowest-order form of closure in a two-mode sense (but should not be thought of as reciprocity because the network is non-directed).

\subsubsection{Longitudinal ERGMs}
\label{sec:longergm}

The development of relational models has primarily focused on cross-sectional data. However, extensions of ERGMs to longitudinal scenarios have been developed -- most often involving a Markov assumption to describe dependence across time. The first longitudinal ERGMs treated tie-formation and tie-dissolution as equitable events in the evolution of the network \cite{Hann:2010}. A more general formulation treats tie-formation (attractiveness in the context of network science) and tie-duration (the complement of tie-duration referred to as fitness in network science) as separable processes, thereby allowing the same network statistic to impact tie-formation and tie-dissolution differently \cite{Kriv:2010}. 

Like ERGMs for cross-sectional data, longitudinal ERGMs are defined by statistics that count the number of occurrences of substructures in the network. However, in addition to the current state of the network, such statistics may also depend on previous states. Under Markovian dependence, network statistics only depend on the current and the most recent state; for example, the number of ties that remain intact from the preceding observation. The recently released tergm (``temporal exponential random graph model'') package in the Statnet suite in R estimates ERGMs for discrete temporal (i.e., longitudinal) sociocentric data \cite{Hann:2010}.

\subsection{Actor-Orientated Approaches}

An alternative approach for modeling network evolution is the actor-oriented model \cite{Snij:1996,Snij:2001,Snij:2005}. This centers on an objective function that actors seek to maximize and which may be sensitive to multiple network properties, including reciprocity, closure, homophily, or contact with high-degree actors. The model assumes that actors control their outgoing ties and change them in order to increase their satisfaction with the network in one or more respects as quantified by the objective function. It resembles a rationale choice model in which each agent attempts to maximize their own utility function. Estimated parameters indicate whether changes in a given property raise or lower actor satisfaction.  

An important distinction of actor-oriented models from ERGMs is that the relevant network statistics in the actor-oriented model are specific to individuals rather than being aggregations across the network. However, like ERGMs, estimation is computationally intensive. The SIENA package in StOCNET \cite{Huis:2004,Huis:2005} uses a stochastic approximation algorithm but struggles with networks of appreciable size (e.g., thousands of individuals). Because they only resemble ERGMs in the limiting steady-state case, actor-oriented models may also suffer from degeneracy but the problem is less profound \cite{Gold:2009}. 

\subsubsection{Joint Models}

A virtue of the actor-oriented modeling framework in SIENA is that an actor's relationships can be modeled jointly with the social-influence effects of an actor's peers on their own traits. If the model is correctly specified, it has the potential to account for unmeasured confounding factors that affect both the evolution of relationship status and the values of individuals attributes, yielding unbiased estimates of the effects of observed variables affecting social influence and the evolution of the network. Such a model was developed by Steglich and colleagues \cite{Steg:2010} but to date work in this area is limited.

\subsection{Latent Independence Approaches}
\label{sec:condindep}

In ERGMs a huge increase in computational complexity occurs between the dyadic independent and dyadic dependent models. A second concern about ERGMs is that in general they are not consistent under sampling in the sense that statistical inferences drawn from the network for the sample do not generalize to the full network \cite{Shal:2012}. The few ERGMs to exhibit such consistency include the dyadic independent p$_{1}$ and stochastic block models. An alternative modeling strategy provides a more graduated transition between independence and dependence scenarios by using random effects to model dyadic dependence and also ensures consistency between the results of analyzing the sample and the population of interest. Random effects are used to account for dyadic independence in the p$_{2}$ model \cite{Duij:2004,Zijl:2006} introduced below.

The p$_{2}$ model is much like the p$_{1}$ model except that the expansiveness $\al_{i}$ and popularity $\ga_{i}$ parameters are random as opposed to fixed effects. Typically, $(\al_{i},\ga_{i})$ is assumed to be bivariate normal with covariance matrix $\bm{\Sigma}_{\al\ga}$. Therefore, the p$_{2}$ model is given by
\begin{equation}
 {\rm pr}(a_{ij},a_{ji} \mid \bm{x}_{ij},\bm{x}_{ji}) = k_{ij}^{-1}\exp(\mu_{ij}a_{ij} + \mu_{ji}a_{ji} + \rho_{ij}a_{ij}a_{ji} ), \label{eq:P2mod}
\end{equation}
\begin{eqnarray*}
 \mbox{where } \kappa_{ij} &=& 1 + \exp(\mu_{ij}) + \exp(\mu_{ji}) + \exp(\mu_{ij} + \mu_{ji} + \rho_{ij}), \\
 \mu_{ij} &=&  \mu + \al_{i} + \ga_{j} + \bm{\be}_{1}^{T}\bm{x}_{ij}, \\
 \rho_{ij} &=& \rho + \bm{\be}_{2}^{T}\bm{x}_{2ij},
\end{eqnarray*}
and $(\al_{i},\ga_{i}) \sim {\rm Normal}(\bm{0},\bm{\Sigma}_{\al\ga})$. Thus, $\bm{x}_{ij}=(\bm{x}_{1ij},\bm{x}_{2ij})$ and $\bm{x}_{2ij}$ includes a subset of covariates that are symmetric ($\bm{x}_{2ij}=\bm{x}_{2ji}$) in reflection of the fact that reciprocity is a symmetric phenomenon. Conditional on $(\al_{i},\ga_{i})$ the model implies that the relationship status of one dyad does not depend on that of another. A positive off-diagonal element of $\bm{\Sigma}_{\al\ga}$ implies that expansive individuals also tend to be popular.

The p$_{2}$ model can be extended to account for more general forms of dyadic dependence than the latent propensity of an individual to send or receive ties. Let each individual have a vector of latent variables, denoted $z_{i}$ in the case of individual $i$, that together with the same for individual $j$ affects the value of the relationship between $i$ and $j$. The dependence of tie-status on $\bm{z}_{i}$ is generally represented using a simple mathematical function. The major types of models are latent class models \cite{Nowi:2001,Airo:2008}, latent distance models \cite{Hoff:2002,Hand:2007}, and latent eigenmodels \cite{Hoff:2005,Hoff:2008}. These models are characterized by the form of the latent variable
\begin{equation}
 \xi(\bm{z}_{i},\bm{z}_{j}) = \left\{ \begin{array}{l} \la_{\bm{z}_{i},\bm{z}_{j}} \mbox{ where $\bm{z}_{i},\bm{z}_{j} \in \{1,\ldots,K\}$ and } \la_{\bm{z}_{i},\bm{z}_{j}}=\la_{\bm{z}_{j},\bm{z}_{i}} \\
 -|\bm{z}_{i}-\bm{z}_{j}|^{c} \mbox{ where $c>0$ and $\bm{z}_{i},\bm{z}_{i}$ have $K$ elements} \\
 \bm{z}_{i}^{T}\bm{U}\bm{z}_{j} \mbox{ where $\bm{z}_{i} \sim N(0,\bm{\Sigma_{z}})$ and $U$ is a $K$-dimensional diagonal matrix} \end{array}, \right. \label{eq:LatVble}
\end{equation}
which is included as an additional predictor in $\mu_{ij}$. In (\ref{eq:LatVble}) the form and interpretation of $\bm{z}_{i}$ changes from denoting a scalar $\xi(\bm{z}_{i},\bm{z}_{j})$ categorical latent variable in the latent class model (first row), to a position in a continuously-valued multi-dimensional space in the latent distance and latent eigenmodels (second and third rows, respectively). The term $\xi(\bm{z}_{i},\bm{z}_{j})$ can be added to either the $\mu_{ij}$ or $\rho_{ij}$ components of the p$_{2}$ model to allow higher-order dependence to moderate the effect of density and reciprocity, respectively.

In the latent class specification the array of values of $\la_{\bm{z}_{i},\bm{z}_{j}}$ form a symmetric $K \times K$ matrix $\bm{\La}$. A basic specification is $\la_{\bm{z}_{i},\bm{z}_{j}}=\la_{0}$ if $\bm{z}_{i} = \bm{z}_{j}$ (nodes in same partition) and $\la_{\bm{z}_{i},\bm{z}_{j}}=0$ if $\bm{z}_{i} \neq \bm{z}_{j}$ (nodes in different partitions) (\citeNP{Nowi:2001,Airo:2008}). Latent class models extend stochastic-block models to allow latent clusters as well as observed clustering variables. This family of models is suited to network data exhibiting structural equivalence; that is, under the model individuals are hypothesized to belong to latent groups such that members of the same group have similar patterns of relationships. 

In the latent distance specification the most common values for $c$ are 1 and 2, corresponding to absolute and cartesian distance, respectively. The distance metric accounts for latent homophily -- the effect of unobserved individual characteristics that induce ties between individuals. In this model, $\bm{z}_{i}$ can be interpreted as the position of individual $i$ in a social space \cite{Hoff:2002}. This model accounts for triadic dependence (e.g., transitivity) by requiring that latent distances between individuals obey the triangle inequality. Latent distance models are available in the LatentNet package in R \cite{Kriv:2008}. 

The latent eigenmodel is the most general specification and accounts for both structural equivalence and latent homophily. Furthermore, the parameter space of the latent eigenmodel model of dimension $K$ generalizes that of the latent class model of the same dimension and weakly generalizes the latent distance model of dimension $K-1$. Conversely, the latent distance model of dimension $K$ does not generalize the one-dimensional latent eigenmodel model \cite{Hoff:2008}. The closeness of the latent factors $\bm{U^{1/2}z}_{i}$ and $\bm{U^{1/2}z}_{j}$ quantifies the structural equivalence of actors $i$ and $j$ positions in the network; a tie is more likely if $\bm{U^{1/2}z}_{i}$ and $\bm{U^{1/2}z}_{j}$ have a similar direction and magnitude, allowing for more clustering than under (\ref{eq:P2mod}). On the other hand, latent homophily is accounted for by the diagonal elements of $\bm{U}$, which can be positive or negative (allowing for heterophily as well as homophily). The model constrains the extent to which the quadratic-forms $\bm{z}_{i}^{T}\bm{Uz}_{j}$, $\bm{z}_{i}^{T}\bm{Uz}_{k}$, and $\bm{z}_{j}^{T}\bm{Uz}_{k}$ constructed from the latent vectors vary from one another. The greater the magnitude of $\bm{\Sigma_{z}}={\rm cov}(\bm{z}_{i})$ the greater the extent to which ties are expected to cluster and form cliques. The latent eigenmodel model is appropriate if a network exhibits clustering due to both structural equivalence and unmeasured homophily.

In \citeN{Hoff:2005} and \citeN{Hoff:2008} models are specified at the tie-level with reciprocity (in directed networks) represented as the within-dyad correlation between two tie-specific latent variables. Modeling reciprocity as a latent process differs from the p$_{2}$ model, in which reciprocity is represented as a direct effect \cite{Paul:2013}. Therefore, an alternative family of latent variable models for networks is obtained by augmenting the density term in the p$_{2}$ model with (\ref{eq:LatVble}). An advantage of specifying a joint model at the dyad level is that the resulting (extended-p$_{2}$) model involves $N(N-1)$ fewer latent variables, possibly alleviating computational issues such as non-identifiability of parameters or multiple local optima.

The challenges of estimating models involving latent variables resemble those of factor analysis or other dimension-reduction methods. First, an appropriate value of $K$ may not be able to be specified from existing knowledge of the network and estimating $K$ from the data is not straightforward. Second, computational challenges in estimating the latent variables can make the method difficult to apply to large networks. However, such issues are more easily overcome than degeneracy in ERGMs. Degeneracy is avoided in these models as the model for a dyad determines the distribution of the network. In other words, the factorization of the likelihood into a product of like terms ensures asymptotically that networks sampled under the model are almost surely in the neighborhood of the observed network, increasingly so as $N$ increases. Another contrast with ERGMs is that the model describes a population as opposed to the single observed network. Thus, in latent variable models the data generating process is modeled whereas ERGMs are specific to the observed network and so have more in common with finite population inference. 

Another advantage of conditional independence models over ERGMs is that the same types of models can be applied to valued relational data. Analogous to generalized linear models, the link function and any parametric distributions assumptions that define a conditional independence network model can be tailored to the type of relationship variable (scale, count, ratio, categorical, multivariate). However, a recent adaptation of ERGMs has been proposed for modeling count-valued sociocentric data \cite{Kriv:2012}. 

Offsetting the above advantageous features of conditional independence models is that terms such as $\xi(\bm{z}_{i},\bm{z}_{j})$ are limited from the hypothesis testing and interpretational standpoint in that they do not distinguish particular forms of social equivalence or latent homophily. For example, the effect of transitivity is not distinguished from that of cyclicity or higher-order clustering, such as tetradic closure. Therefore, the choice of model in practice might depend on the importance of testing specific hypotheses about higher-order effects to obtaining a model whose generative basis allows it to make predictions beyond the data set on which the model was estimated.

\subsubsection{Longitudinal conditional independence models}
\label{sec:longit}

Longitudinal counterparts of conditional independence models are obtained by introducing terms that account for longitudinal dependence (e.g., past states of the dyad). A simple Markov transition model was developed in \citeN{Omal:2011} with tie-formation and tie-dissolution treated as unrelated processes. Conditional on the past state of the dyad and the sender and receiver random effects, the value of each tie is assumed to be statistically independent of that of any other tie. A more general formulation extends the p$_{2}$ model, allowing dependence between ties within a dyad (reciprocity), heterogeneous effects in the formation and dissolution of ties, and the inclusion of higher-order effects (e.g., third-order interactions to account for transitivity) as lagged predictors \cite{Paul:2013}. 

The approach in \citeN{Paul:2013} is notable for attempting to capture the best of both worlds: it allows localized (actor or dyadic) versions of the higher-order predictors available in ERGMs to be included as predictors, but avoids degeneracy by using their lagged values as opposed to their current values as predictors. Therefore, conditional on the observed and latent predictors, dyads are cross-sectionally independent but longitudinally dependent on prior states of other dyads (in addition to their own past states) in the network. An extension that builds on \citeN{Paul:2013} is to incorporate the latent class, distance or eigenfactor terms in (\ref{eq:LatVble}) in the model. Such a model was entertained in \citeN{West:2011} but has not yet been developed.

\section*{Part III: Network Science}

We now switch attention to methods that have been derived and used in the field of network science. In general, network science approaches avoid assumptions about distributions in models. For example, to test whether a network exhibits a certain property, the commonly-employed approach is to use a permutation test to develop a null distribution for a statistic that embodies the property in question and then evaluate how extreme the observed value of the statistic is with respect to the null distribution. This technique is the cornerstone of the procedure used to evaluate the degree of separation to which social clustering can be detected in \citeN{Szab:2007}. 

Network science focuses not only on social networks, but also covers information networks, transportation networks, biological networks, and many others. Most of the networks studied within network science are non-directed as ties are typically thought of as connections as opposed to measures for which the distinction between instigator and receiver is relevant. Thus, the networks in this section are assumed to be non-directed unless stated otherwise. 

\section{Generative Models of Network Formation}

Network science has taken a somewhat different approach to modeling networks than the social sciences or statistics. Essentially all models developed within network science are {\it generative models}, sometimes also known as forward models, in contrast to probabilistic models such as ERGMs. These models start from a set of simple hypothesized mechanisms, often functioning at the level of individual nodes and ties, and attempt to describe what types of network structures emerge from a repeated application of the proposed mechanisms. Many of the models describe growing networks, where one starts from a small connected seed network consisting of a few connected nodes, and then grows the network by subsequent addition of nodes, usually one at a time. The \emph{attachment rules} specify how exactly an incoming node attaches itself to the existing network.

Generative models are commonly exploratory in nature. If they reproduce the type of structure observed in an empirical network, it is plausible that the proposed mechanisms may underlie network formation in the real world. The main insight to be gained from a generative model is a potential explanation for why a network possesses the type of structure it does. Many of the models are simple in nature, which occasionally leads to analytical tractability, but the main reason for simplicity is the potential to expose clearly the main mechanism(s) driving the phenomenon of interest. It is not uncommon for generative models to possess only two or three parameters, yet occasionally simple generative mechanisms can explain some of the key features surprisingly well. Once a model can explain the main features, it can be fine-tuned by adding more specific or nuanced mechanisms. A few examples of generative models are now described.

\subsection{Cumulative Advantage Model}

Cumulative advantage refers to phenomena where success seems to breed success, such as in the case of  accumulation of further wealth to already wealthy individuals. In networks of scientific citations, where a node represents a scientific paper, each node has some number of edges pointing to nodes that correspond to cited papers \cite{price1965}. In the present context, for example, there would be an edge pointing from the node representing this chapter to the node representing the 1965 \emph{Science} paper of Price. While the out-degree of nodes is fairly uniform, as the length of bibliographies is fairly constrained, the in-degree distribution was found to be fat-tailed with the functional form of a power-law, $P(k) \sim k^{-\alpha}$ \cite{price1965}.

Price later proposed a mathematical model for cumulative advantage processes, ``the situation in which success breeds success'' \cite{price1976}. In this model, nodes are added to the network one at a time, and the average out-degree of each node is fixed. The attachment rule in the model specifies that each new paper will cite existing papers with probability proportional to the number of citations they already have. Thus each incoming node will attach itself with some number of directed edges to the existing network, the exact number of ties being drawn from a distribution, and the nodes these new edges are pointing to will be chosen proportional to their in-degree. In this formulation, however, papers with exactly zero citations can never accrue  citations. To overcome this problem, one can either consider the original publication as the first citation so that each paper starts with one citation or, alternatively, add a small constant to the number of citations \cite{price1976}. Either way, the outcome is that the target nodes are chosen in proportion to their in-degree plus this small positive constant. A derivation of the resulting in-degree distribution is given by Newman \cite{newman2010}. Denoting the average out-degree of a node by $c$ and using $a$ to denote the small positive constant, the in-degree distribution $P(k)$ for large values of $k$ has the power-law form $P(k) \sim k^{-\alpha}$, where $\alpha = 2 + a/c$.

This simple model (although the derivation of the result is quite involved) is able to reproduce the empirical citation (in-degree) distribution for scientific papers with surprising accuracy given that the model only contains two parameters. It may seem odd that the model does not incorporate any notion of paper quality, which surely should be an important driver of citations. Here it is important to notice that the model does not make any attempt to predict \emph{which} paper becomes popular (although it can be shown, using the model, that papers published at the inception of a field have a much higher probability to become popular). Instead, the model incorporates the quality of papers implicitly, and indeed the number of citations to a paper is frequently seen as an indicator of its quality. Popular papers are also easily discovered, which further feeds their popularity. The idea of using popularity as a proxy for quality may extend to other areas where resources are scarce; for example, skilled surgeons are in high demand.

\subsection{Preferential Attachment Model}

The cumulative advantage model of Price \cite{price1976} is developed as a modification of the Polya urn model, which is used to model a sampling process where each draw from the urn, corresponding to a collection of different types of objects, changes the composition of the urn and thereby changes the probability of drawing an object of any type in the future. The standard Polya urn model consists of an urn containing some number of black and white balls, drawing a ball at random and then returning it to the urn along with a new ball of the same color \cite{feller1966}. Independently of Price, Barabasi and Albert introduced a similar model in 1999 \cite{barabasi1999}. They examined the degree distributions of an actor collaboration network (two actors are connected if they are cast in the same movie), World Wide Web (two web pages are connected if there is a hyperlink from one page to the other), and power grid (two elements (generators, transformers, substations) are connected if there is a high-voltage transmission line between them), finding that they approximately followed power-law distributions. Although the actor collaboration network and the power grid networks are defined much like a projection from a two-mode to a one mode network, a subtle difference between them is that direct interaction between the nodes can be assumed. In other words, the nodes can be thought of as directly linked. 

Both of the generic network models in existence at the time, the \er and the \ws models, operated on a fixed set of $N$ vertices, and assumed that connections were placed or rewired without any regard to the degrees of the nodes to which they were connected. The model of Barabasi and Albert changed both of these aspects. First, they introduced the notion of network growth, such that at each time step a new node would be added to the network. Second, this new node would connect to the existing network with exactly $m$ non-directed edges, and the nodes they attached to were chosen in proportion to their degree. The probability for the incoming vertex to connect to vertex $i$ depends solely on its degree $k_i$ and is given by
$$\Pi(k_i) = k_i / \sum_{j}k_j.$$

The model was solved by Barabasi and Albert using rate equations, which are differential equations for the evolution of node degree over time where both degree and time, as an approximation, are treated as if they were continuous variables \cite{barabasi1999,barabasi1999a}. More general solutions were provided by Krapivsky \emph{et al.} also using rate equations \cite{krapivsky2000} and Dorogovtsev \emph{et al.} using master equations which, like rate equations, are differential equations for the evolution of node degree, but they (correctly) treat degree as a discrete variable while still making the continuous-time approximation for time \cite{dorogovtsev2000}. In the master equation approach, one writes down an equation for the evolution of the number of nodes of a given degree. Let us use $N_k(t)$ to denote the number of nodes of degree $k$ in the network at time $t$, where time is identified with network size, i.e., time $t$ corresponds to the network at the point of its evolution when it consists of $t$ nodes. (The nodes making up the seed network can be usually ignored in the limit as time increases.) The number $N_k(t)$ can change in two ways: it can either increase as an incoming node attaches itself to a node of degree $k-1$ and thus turns it into a node of degree $k$, or it can decrease as an incoming node attaches itself to a node of degree $k$, turning into a node of degree $k+1$. The former situation leads to $N_k(t+1) = N_k(t) + 1$, and the latter to $N_k(t+1) = N_k(t) - 1$. Transitions larger than one, e.g., from $k$ to $k+2$ or from $k$ to $k-2$ are very unlikely and can be ignored. The value of $N_m(t)$ increases by one per time step as each incoming node has degree $m$, which also means there are no nodes with degree less than $m$, and hence the equations used to model the evolution of quantities like $N_k(t)$ are not valid for $k < m$. The resulting degee distribution has the form
\[ P(k) = \frac{2m(m+1)}{k(k+1)(k+2)}, \] which asymptotically behaves as $P(k) \sim k^{-3}$.

The preferential attachment model of Barabasi and Albert has attracted a tremendous amount of scientific interest in the past few years, and consequently numerous modifications of the model have been introduced. For example, extensions of the model allow:
\begin{itemize}
\item Ties to appear and disappear between any pairs of vertices (the original formulation only considers the addition of ties between the incoming vertex and set of vertices already in existence).
\item Vertices to be deleted either uniformly at random or based on their connectivity.
\item The attachment probability $\Pi(k_i)$ to be super-linear or sub-linear in degree, or to consist of several terms.
\item Nodal attributes, such as the {\it attractiveness} (the propensity with which new ties form with the node) or {\it fitness} (the propensity with which established ties remain intact) of a node, and the attachment probability can incorporate these attributes in addition to degree.
\item Edges to assume weights instead of $\{0,1\}$ binary values to codify connection strength between any pair of elements. 
\end{itemize}
In the context of physician networks, a preferential attachment model could be used to examine the process of new physicians seeking colleagues to ask for advice upon joining a medical organization, such as a hospital. Under the preferential attachment hypothesis, new physicians would be more likely to form ties with and thus seek advice from popular established physicians or physicians in the same cohort (e.g., Medical school or residency program).

\subsection{Social Network Models}

The class of models known as \emph{network evolution models} can be defined via three properties: (i) the models incorporate a set of stochastic attachment rules which determine the evolution of the network structure explicitly on a time-step--by--time-step basis; (ii) the network evolution starts from an empty network consisting of nodes only, or from a small seed network possessing arbitrary structure; and (iii) the models incorporate a stopping criterion, which for growing network models is typically in the form of the network size reaching a predetermined value, and for dynamical (non-growing) network models the convergence of network statistics to their asymptotic values. Many network evolution models do not reference intrinsic properties or attributes of nodes, and in this sense they are similar to the various implementations of preferential attachment models that do not postulate node specific fitness or attractiveness. 

Most network evolution models that are intended to model social networks employ some variants of focal closure and cyclic closure (see, e.g., \citeN{kossinets2006}). \emph{Focal closure} refers to the formation of ties between individuals based on shared foci, which in a medical context could correspond to a group of doctors who practice in a particular hospital (the focus). The concept of shared foci in network science is analogous to homophily in social network analysis. More broadly, ties could represent any interest or activity that connects otherwise unlinked individuals. In contrast, \emph{cyclic closure} refers to the idea of forming new ties by navigating and leveraging one's existing social ties, a process that results in a cycle in the underlying network. Because the network is non-directed, the term cycle is used interchangeably with closure. This differs from when the network is directional and a cycle is a specific form of closure, with transitivity being another form. \emph{Triadic closure}, which is the special case of cyclic closure involving just three individuals, refers to the process of getting to know friends of friends, leading to the formation of a closed triad in the non-directed network. Most social networks are expected to (i) have skewed and fat-tailed degree distributions, (ii) be assortatively mixed (high-degree individuals are connected to high-degree individuals), (iii) be highly clustered, and (iv) possess the small-world property (average shortest path lengths are short, or more precisely, scale as $\log(N)$), and (v) exhibit community structure.

The models by Davidsen \emph{et al.} \cite{davidsen2002} and Marsili \emph{et al.} \cite{marsili2004} exemplify dynamic (non-growing) network evolution models for social networks. Both have a mechanisms that starts by selecting a node $i$ in the network uniformly at random. In the model of Davidsen \emph{et al.}, if node $i$ has fewer than two connections, it is connected to a randomly chosen node in the network; otherwise two randomly chosen neighbors of node $i$ are connected together. In the model of Marsili \emph{et al.}, node $i$ (regardless of its degree) is connected with probability $\eta$ to a randomly chosen node in the network; then a second-order neighbor of node $i$, i.e., a friend's friend, is connected with probability $\xi$ to node $i$. The first mechanism in each model, the random connection, emulates focal closure, because there are no nodal attributes signifying shared interests. The point is that the formation of these connections is not driven by the structure of existing connections but, from the point of view of network structure, is purely random. The second mechanism, the notion of triadic closure, is implemented in slightly different ways across the models. If these mechanisms were applied indefinitely, the result would be a fully connected network. To avoid this outcome, the models also delete ties at a constant rate, which makes it possible for network statistics of interest to reach stationary distributions. In the model of Davidsen \emph{et al.}, tie deletion is accomplished by choosing a node in the network uniformly at random, and then removing all of its ties with some probability; Marsili \emph{et al.} accomplish the same phenomenon by selecting a tie uniformly at random, and then deleting it with probability $\lambda$. Growing network evolution models, such as those by V\'azquez \cite{vazquez2003} and Toivonen \emph{et al.} \cite{toivonen2006}, do not usually incorporate link deletion, but instead grow the network to a pre-specified size, which obviates the need for link deletion.

Marsili \emph{et al.} use extensive numerical simulations, as well as a master equation approach applied to a mean-field approximation of the model, to explore the impact of varying the probabilities $\eta$ (global linking), $\xi$ (neighborhood linking), and $\lambda$ (link deletion) for average degree and average clustering coefficient. Consider a situation where the value of $\xi$ (neighborhood linking) is increased while keeping the value of $\lambda$ (link deletion) fixed. At first, for small values of $\xi$, components with more than two nodes are rare, and the network can be said to be in the sparse phase. Upon increasing the value of $\xi$ up to a specific point, a large connected component emerges, and the value of the average degree suddenly jumps up. This point equals $\xi_2 / \lambda$ and is known as the critical point -- it marks the beginning of the dense phase in the phase diagram of the system. As $\xi$ is increased further, the network becomes more densely connected. Reversing the process by slowly decreasing the value of $\xi$ identifies a range of values from $\xi_1 \le \xi \le \xi_2$ where the largest connected component remains densely connected and the average degree remains high. Only when the value of $\xi$ is decreased below a point denoted by $\xi_1$ does the network ``collapse'' and re-enter the sparse phase. This phenomenon, which demonstrates some of the connections between network science and statistical physics, is typical of first-order or discontinuous phase transitions in statistical physics, and it demonstrates how hysteresis, the effect of the system remembering its past state can rise in networked systems. Although Markov dependence is a special case of hysteresis, its use is generally restricted to probabilistic models whereas hysteresis is typically aligned with nonlinear models of physical phenomena having a continuous state-space. From the social network point of view this means that the network can remain in a connected phase even if the rate of establishing new connections at the current rate would not be sufficient for getting the network to that phase in the first place. In more practical terms, this finding implies that it is possible to maintain a highly connected network with a relatively low ``effort'' (the $\xi$ parameter in the model) once the network has been established, but that same low level of effort would not be sufficient for establishing the dense phase of network evolution in the first place. (The analogy in social network analysis is that the threshold for forming a (e.g.) friendship is greater than that needed for it to remain intact.)

\begin{figure}[htbp]
\begin{center}
\includegraphics[width=1\linewidth]{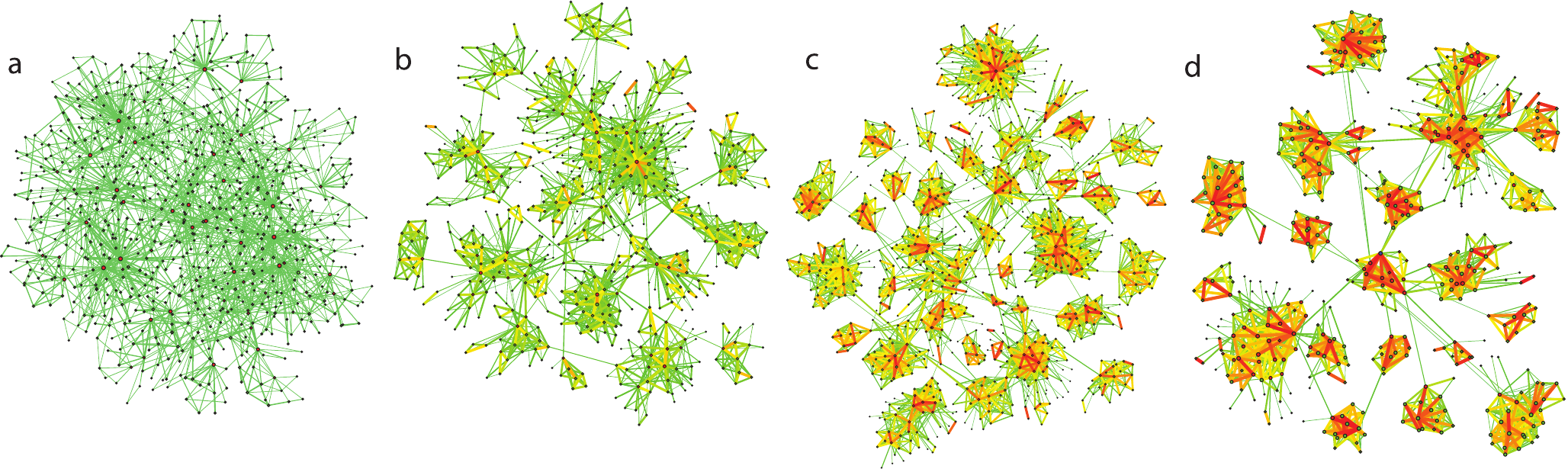}
\caption{Network structures produced by the model of Kumpula \emph{et al.} by varying the reinforcement parameter as follows: (a)  $\de = 0$, (b)  $\de = 0.1$, (c)  $\de = 0.5$, and (d) $\de = 1$. Figure adapted from Kumpula {\it et al.} (2007).}
\label{fig:kumpula}
\end{center}
\end{figure}

The model by Kumpula \emph{et al.} \cite{kumpula2007}, which is another dynamical (non-growing) network evolution model for social networks, implements cyclic closure and focal closure (see Figure~\ref{fig:kumpula}) in a manner similar to the models of Davidsen \emph{et al.} and Marsili \emph{et al.}, but introduces a minor modification.

Unlike the previous models which produce binary networks with $A_{ij} = \{0,1\}$, this model produced weighted networks with $A_{ij} \ge 0$. The main modification deals with the triadic closure step, which here is implemented as a weighted two-step random walk. Starting from a randomly chosen node $i$; this node chooses one of its neighbors $j$ with probability $w_{ij} / s_i$, where $s_i = \sum_{j} w_{ij}$ is the strength of node $i$, i.e., the sum of the edge weights connecting it to its neighbors. If node $j$ has neighbors other than $i$, such a node $k$ will be chosen with probability $w_{jk} / (s_j - w_{ij})$, where there is a requirement that $k \ne i$. The weights $w_{ij}$ and $w_{jk}$ on the edges just traversed will be increased by a value $\de$. In addition, if there is a link connecting node $i$ and node $k$, the weight $w_{ik}$ on that link is similarly increased by $\de$; otherwise a new link is established between node $i$ and $k$ with $w_{ik} = 1$. When $\de = 0$, there is no clear community structure present, but as the value of $\de$ is increased, very clear nucleation of communities takes place. This phenomenon happens because when $\de > 0$, a type of positive feedback or memory gets imprinted on the network, which reinforces existing connections, and makes future traversal of those connections more likely. This is not unlike the models of cumulative advantage or preferential attachment discussed above, but now applies to individual links as opposed to nodes. If one inspects the community structure produced by the model, most of the strong links appear to be located within communities, whereas links between communities are typically weak. This type of structural organization is compliant with the so-called weak ties hypothesis, formulated in \citeN{granovetter1973}, which states, in essence, that the stronger the tie connecting two individuals, the higher the fraction of friends they have in common. Onnela \emph{et al.} showed that a large-scale social network constructed from the cell phone communication records of millions of people was in remarkable agreement with the hypothesis -- only the top 5\% of ties in terms of their weight deviated noticeably from the prediction. The networks produced by the model of Kumpula \emph{et al.} are clearly reminiscent of observed real-world social networks, and the inclusion of the tuning parameter $\de$ makes it straightforward to create networks with sparser or denser communities. The downside is that the addition of weights to the model appears to make it analytically intractable.

Nodal attribute models, in stark contrast to network evolution models, specify nodal attributes for each node, which could be scalar or vector valued. The probability of linkage between any two nodes is typically an increasing function of the similarity of the nodal attributes of the two nodes in consideration. This is compatible with the notion of homophily, the tendency for like to attract like. Nodal attribute  models can also be interpreted as spatial models, where the idea is that each node has a specific location in a social space. The models by Bogu\~n\'a \emph{et al.} \cite{boguna2004} and Wong \emph{et al.} \cite{wong2006} serve as interesting examples. Nodal attribute models do not specify attachment rules at the level of the network, and in some sense can be seen as latent variable models for social network formation. These types of models have been studied less in the network science literature than network evolution models. 

Clearly, nodal attribute models have a strong resemblance to models developed and studied in the social network literature that treat dyads as independent conditional on observed attributes of the individuals, other covariates, and various latent variables (individual-specific random effects in the case of the p$_{2}$ model, categorical latent variables in the case of latent class models, continuous latent variables under the latent-space and latent eigenmodels in Section~\ref{sec:condindep}). Unlike network science, work on such models in the social network literature has been more prominent than work on network evolution. A difference in the approach of some nodal attribute models and social network models is that the former may use specific rules for determining whether a tie is expected, such as a threshold function (in a sense emulating formal decision making), whereas the latter rewards values of parameters that make the model most consistent with the observed network(s).

\section{Network Communities}
\label{sec:PhysCD}

Many network characteristics are either microscopic or macroscopic in nature; the value of a microscopic characteristic depends on local network structure only, whereas the value of a macroscopic characteristic depends on the structure of the entire network. Node degree is an example of a microscopic quantity: the degree of a node depends only on the number of its connections. In contrast, network diameter, the longest of all pairwise shortest paths in the network, can change dramatically by the addition (or removal) of even a very small number of links anywhere in the network. For example, a $k$-cycle consists of $k$ nodes connected by $k$ links such that a cycle is formed with each node connected to precisely two nodes. The diameter of such a network is $\lfloor k/2 \rfloor$, where the floor function $\lfloor x \rfloor$ maps a real number $x$ to the largest previous integer, such that for an even $n$ it follows that $\lfloor k/2 \rfloor = k/2$. For large values of $k$, adding just a few links quickly brings down the value of network diameter. There is a third, intermediate scale that lies between the microscopic and macroscopic scales which is often known as the \emph{mesoscopic} scale. For example, a $k$-clique could justifiably be called a mesoscopic object (especially if $k$ is large). Another type of mesoscopic structure is that of a network community, which can be loosely defined as a set of nodes that are densely connected to each other but sparsely connected to other nodes in the network (but not to the extent of resulting in distinct components).

\begin{figure}[htbp]
\begin{center}
\includegraphics[width=0.5\linewidth]{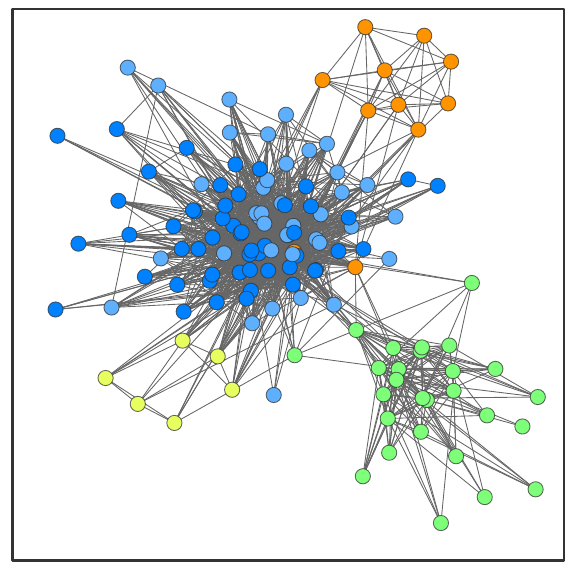}
\caption{Communities in a patient-sharing network of physicians. Each vertex corresponds to a physician, and a pair of physicians are connected with a tie if they share patients. The community assignment of each physician is indicated by the node color. In this case the ``green'' and ``orange'' communities are fairly distinct.}
\label{fig:commex}
\end{center}
\end{figure}

There has been considerable interest especially in the physics literature focusing on how to define and detect such communities, and several review papers cover the existing methods \cite{porter2009,fortunato2010,newman2012}. The motivation behind many of these efforts is the idea that communities may correspond to functional units in networks, such as unobserved societal structures. The examples range from metabolic circuits within cells \cite{guimera2005} to tightly-knit groups of individuals in social networks \cite{newman2004,traud2012}. The interested reader can consult the review articles on community detection methods \cite{porter2009,fortunato2010,newman2012} for more details. Another application is health care where, for instance, \shortciteN{Land:2012} have deduced communities of physicians based on network ties representing their treating the same patients within the same period of time. The clustering of physicians in communities is shown for one particular Hospital Referral Region (a health care market encompassing at least one major city where both cardiovascular surgical procedures and neurosurgery are performed) in the United States (Figure~\ref{fig:commex}). 

One potential application of network science methods for community detection is in the area of health education and disease prevention (e.g., screening). Due to limited resources, it may not be possible to send materials or otherwise directly educate every member of the population. The partition of individuals into groups would facilitate a possibly more efficient approach whereby the communities are first studied to identify key individuals. Then a few key individuals in each community are trained and advised on mechanisms for helping the intervention to diffuse across the community. A general characteristic of interventions where such an approach might be useful are those where intensive training is required to be effective and where delegation of resources through passing on knowledge or advice is possible.

\subsection{Modularity maximization}

A number of network community detection methods define communities implicitly via an appropriately chosen quality function. The underlying idea is that a given network can be divided into a large number of partitions, or subsets of nodes, such that each node belongs to one subset, and each such partition $\mathcal{P}$ has a scalar-valued quality measure associated with it, denoted by $Q(\mathcal{P})$. In principle one would like to enumerate all possible partitions and compute the value of $Q$ for each of them, and the network communities would then be identified as the partition (or possibly partitions) with the highest quality. In practice, however, the number of possible partitions is exceedingly large even for relatively small networks, and therefore heuristics are needed to optimize the value of $Q$. Community detection methods based on quality function optimization therefore have two distinct components, which are the functional form of the quality function $Q$, and the heuristic used for navigating a subset of partitions over which $Q$ is maximized.

The most commonly used optimization based approach to community detection is modularity maximization, where modularity is one possible choice for the quality function $Q$; in statistical terminology, modularity maximization would be regarded as a non-parametric procedure due to the fact that no distributional nor functional form assumptions are relied upon. There are many variants of modularity, but here the focus is on the original formulation by Newman and Girvan \cite{newman2003,newman2004,newman2006}. Modularity can be seen as a measure that characterizes the extent of homophily or assortative mixing by class membership, and one way to derive it is by considering the observed and expected numbers of connections between vertices of given classes, where the class of vertex $i$ is given by $c_i$. The following derivation follows closely that of \cite{newman2010}, although other derivations, based for example on dynamic processes, are also available.

We start by considering the observed number of edges between vertices of the same class, which is given by $\frac{1}{2} \sum_{i,j} A_{ij} \delta(c_i, c_j)$, where $\delta(\cdot, \cdot)$ is the Kronecker delta, and the factor $1/2$ prevents double-counting vertex pairs. To obtain the expected number of edges between vertices of the same class, cut every edge in half, resulting in two stubs per edge, and then connect these stubs at random. For a network with $m$ edges, there are a total of $2m$ such stubs. Consider one of the $k_i$ stubs connected to vertex $i$. This particular stub will be connected at random to vertex $j$ of degree $k_j$ with probability $k_j / 2m$, and since vertex $i$ has $k_i$ such stubs, the number of expected edges between vertices $i$ and $j$ is $k_i k_j / 2m$. The expected number of edges falling between vertices of the same class is now $\frac{1}{2} \sum_{i,j} \frac{k_i k_j}{2m} \delta(c_i, c_j)$. The difference between the observed and expected number of within class ties is therefore $\frac{1}{2} \sum_{i,j} \left( A_{ij} - \frac{k_i k_j}{2m} \right) \delta(c_i, c_j)$. Given that the number of edges varies from one network to the next, it is convenient to deal with the fraction of edges as opposed to the number of edges, which is easily obtained by dividing the expression by $m$, resulting in  
$$Q_{\textrm{M}}(\mathcal{P}) = \frac{1}{2m} \sum_{i,j} \left( A_{ij} - \frac{k_i k_j}{2m} \right) \delta(c_i, c_j). \label{eq:ModMax} $$
The assignment $\mathcal{P}$ of nodes into classes that maximizes modularity $Q_{\textrm{M}}(\mathcal{P})$ is taken as the optimal partition and identifies the assignment of nodes into network communities. Note that modularity can be easily generalized from binary networks to weighted networks, in which case $k_i$ stands for the strength (sum of all adjacent edge weights) of node $i$, and $m$ is the total weight of the edges in the network.

The expression for modularity has an interesting connection to spin models in statistical physics. In a so-called infinite range $q$-state Potts model, each of the $N$ particles can be in one of $q$ states called spins, and the interaction energy between particles $i$ and $j$ is $-J_{ij}$ if they are in the same state and zero if they are not in different states. The energy function of the system, known as its Hamiltonian, is given by the sum over all of the pairwise interaction energies in the system
$$H(\{\sigma\}) = - \sum_{ij} J_{ij} \delta(\sigma_i, \sigma_j),$$
where $\sigma_l$ indicates the spin of particle $l$ and $\{\sigma\}$ denotes the configuration of all $N$ spins. Finding the minimum energy state (the ground state) of the system corresponds to finding $\{\sigma\}$ such that $H(\{\sigma\})$ is minimized. The states of the particles (spins) correspond to community assignments of nodes in the network problem, and minimizing $H(\{\sigma\})$ is mathematically identical to maximizing modularity $Q_{\textrm{M}}(\mathcal{P})$. In the physical system, depending on the interaction energies, the spins seek to align with other spins (interact ferromagnetically) or they seek to have different orientations (interact antiferromagnetically). In the community detection problem, two nodes seek to be in the same community if they are connected by an edge that is stronger than expected; otherwise they seek to be in different communities. This correspondence between the two problems has enabled the application of computational techniques developed for the study of spin systems and other physical systems to be applied to modularity optimization and, more broadly, to the optimization of other quality functions. Simulated annealing, greedy algorithms, and spectral methods serve as examples of these methods. More details and references are available in community detection review articles \cite{porter2009,fortunato2010}.

\begin{figure}[htbp]
\begin{center}
\includegraphics[width=0.7\linewidth]{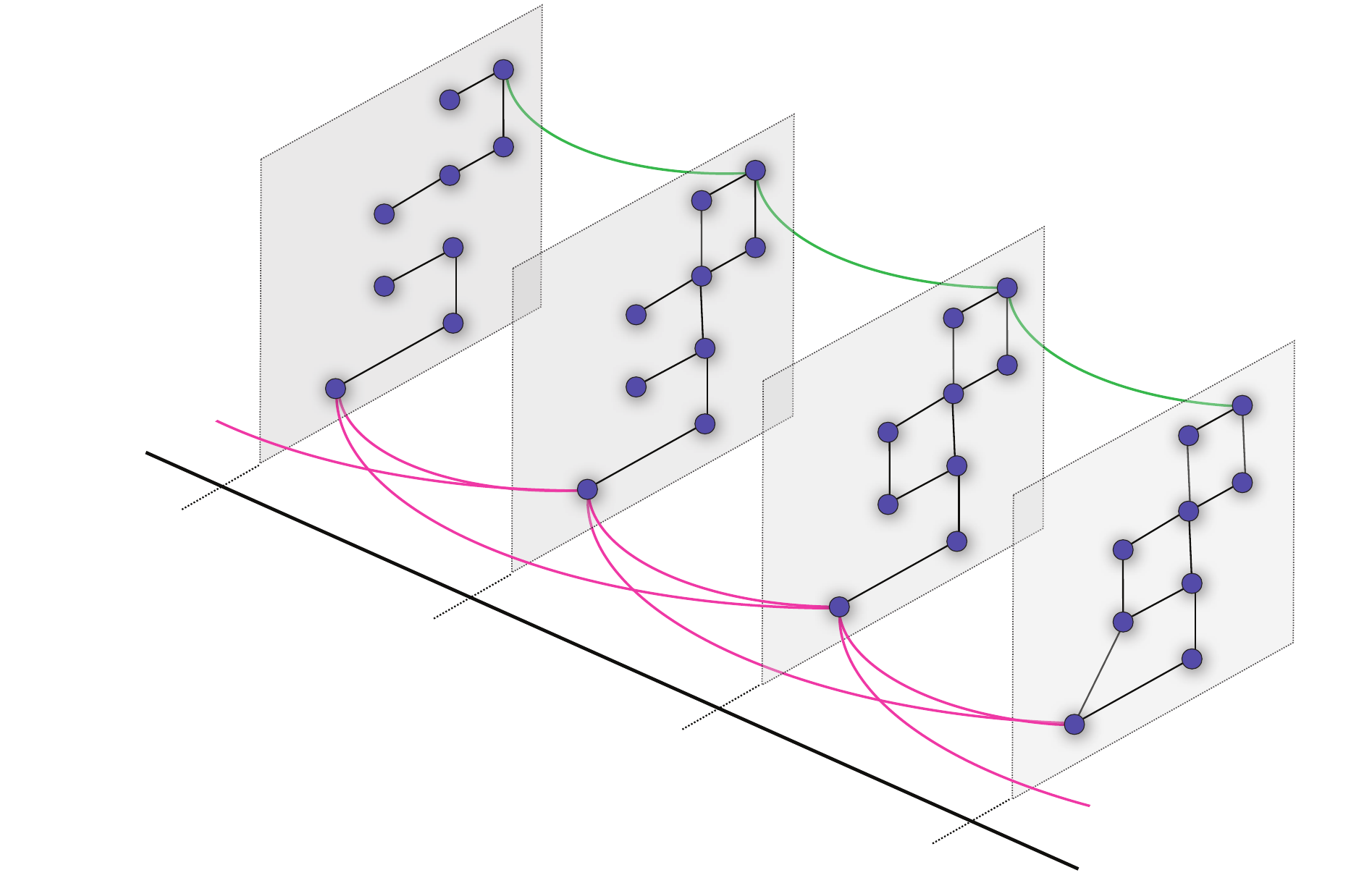}
\caption{Schematic of a multislice network. Each slice represents a network encoded by the adjacency tensor $A_{ijs}$, where subscripts $i$ and $j$ are used to index the nodes and subscript $s$ is used to index the slices. Each node is coupled to itself in the other slices, and the structure of this coupling, encoded by the $C_{jrs}$ tensor, depends on whether the slices correspond to snapshots taken at different times (time-dependent network), to communities detected at different resolution levels (multiscale network), or to a network consisting of multiple types of interactions (multiplex network). For time-dependent and multiscale networks, the slice-to-slice coupling extends for each node a tie to itself across neighboring slices only as exemplified for the node in the upper right corner of the slices; for multiplex networks, the slice-to-slice coupling extends a tie from each node to itself in all the slices as exemplified for the node in the lower left corner. Whatever the form of this coupling, it is applied the same way to each node, although for visual clarity the slice-to-slice couplings are shown just for two nodes. Adapted from Mucha {\it et al.} (2010).}
\label{fig:ms}
\end{center}
\end{figure}

Although there are several extensions of modularity maximization, only one such generalization is described here. Mucha \emph{et al.} developed a generalized framework of network quality functions that allow the study of community structure of arbitrary multislice networks (see Fig.~\ref{fig:ms}), which are combinations of individual networks coupled through links that connect each node in one slice to the same node in other slices \shortcite{mucha2010}. This framework allows studies of community structure in time-dependent, multiscale, and multiplex networks. Much of the work in the area of community detection is motivated by the observation that the behavior of dynamical processes on networks is driven or constrained by their community structure. The approach of Mucha \emph{et al.} is based on a reversal of this logic, and it introduces a dynamical process on the network, and the behavior of the dynamical process is used to identify the (structural) communities. The outcome is a quality function
$$Q_{\textrm{MS}}(\mathcal{P}) = \frac{1}{2 \mu} \sum_{i,j,s,r}\left[ \left( A_{ijs} - \gamma_s \frac{k_{is} k_{js}}{2m_s} \right ) \delta_{sr} + \delta_{ij} C_{jsr} \right ] \delta(c_{is}, c_{jr}),$$
where $A_{ijs}$ encodes the node-to-node couplings within slices and $C_{jrs}$ encodes the node-to-node couplings across slices that are usually set to a uniform value $\omega$; $m_s$ is the number (or weight) of ties within slice $s$ and $\mu$ is the weight of all ties in the network, both those located within slices and those placed across slices; $\gamma_s$ is a resolution parameter that controls the scale of community slices separately for each slice $s$. The standard modularity quality function uses $c_i$ to denote the community assignment of node $i$, but in the multislice context two indices are needed, giving rise to the $c_{is}$ terms, where the subscript $i$ is used to index the node in question and the subscript $s$ to index the slice. The outcome of minimizing $Q_{\textrm{MS}}$, which can be done with the same heuristics as minimization of the standard modularity $Q_{\textrm{M}}$, is a matrix $\mathbf{C}$ that consists of the community assignments $c_{is}$ of each node in every slice.

The multislice framework can handle any combination of time-dependent, multiscale, and multiplex networks. For example, the slices in Fig.~\ref{fig:ms} could correspond, say, to a longitudinal friendship network of a cohort of college students, each slice capturing the offline friendships of the students in each year. If data on the online friendships of the students were also available, corresponding to a different type of friendship, one could then introduce a second stack of four slices encoding those friendships. The four offline slices and the four online slices form a multiplex system, and they would be coupled accordingly. One could further introduce multiple resolution scales, and if one was interested in examining the community structure of the students at three different scales using, say, $\gamma_s \in \{0.5, 1, 2\}$, this would result in a three-fold replication of the $4 \times 2$ slice array with each of the three layers having a distinct value for $\gamma_s$. Taken together, this would lead to a three-dimensional $4 \times 2 \times 3$ array of slices.

\subsection{Clique percolation}

Cliques are (usually small) fully connected subgraphs, and a non-directed $k$-clique is a complete subgraph consisting of $k$ nodes connected with $k(k-1)/2$ links. In materials science the term percolation refers to the movement of fluid through porous materials. However, in mathematics and statistical physics, the field of percolation theory considers the properties of clusters on regular lattices or random networks, where each edge may be either open or closed, and the clusters correspond to groups of adjacent nodes that are connected by open edges. The system is said to percolate in the limit of infinite system size if the largest component, held together by open edges, occupies a finite fraction of the nodes. The method of $k$-clique percolation in \shortciteN{palla2005} combines cliques and percolation theory, and it relies on the empirical observation that network communities seem to consist of several small cliques that share many of their nodes with other cliques in the same community. In this framework, cliques can be thought of as the building blocks of communities. A $k$-clique community is then defined as the union of all adjacent $k$-cliques, where two $k$-cliques are defined to be adjacent if they share $k-1$ nodes. One can also think about ``rolling'' a $k$-clique template from any $k$-clique in the graph to any adjacent $k$-clique by relocating one of its nodes and keeping the other $k-1$ nodes fixed. A community, defined through the percolation of such a template, then consists of the union of all subgraphs that can be fully explored by rolling a $k$-clique template. As $k$ becomes larger, the notion of a community becomes more stringent, and values of $k=3,\ldots,6$ tend to be most appropriate because larger values become unwieldy. The special case of $k = 2$ reduces to bond (link) percolation and $k = 1$ reduces to site (node) percolation.
 
The $k$-clique percolation algorithm is an example of a local community-finding method. One obtains a network's global community structure by considering the ensemble of communities obtained by looping over all of its $k$-cliques. Some nodes might not belong to any community (because they are never part of any $k$-clique), and others can belong to several communities (if they are located at the interface between two or more communities).  The nested nature of communities is recovered by considering different values of $k$, although $k$-clique percolation can be too rigid because focusing on cliques typically causes one to overlook other dense modules that are not quite as tightly connected.  

The advantage of $k$-clique percolation is that it provides a successful way to consider community overlap. Allowing the detection of network communities that overlap is especially appealing in the social sciences, as people may belong simultaneously to several communities (colleagues, family, friends, etc.). However, the case can be made that it is the underlying interactions that are different, and one should not combine interactions that are of fundamentally different types. In statistics, this is analogous to using composite variables or scales that combine multiple items in (e.g.) health surveys or questionnaires. If the nature of the interactions is known, the system might be more appropriately described as a multiplex network, where one tie type encodes professional interactions, another tie type corresponds to personal friendships, and a third tie type captures family memberships. The multislice framework discussed above is able to accommodate memberships in multiple communities as long as distinct interaction types are encoded with distinct (multiplex) ties.

\subsection{Comparison to Social Network approaches to "Community Detection"}

The latent class models in Section~\ref{sec:condindep} partitions the actors in a network into disjoint groups that can be thought of as communities. The clustering process can be thought of as a search for structural equivalence in that individuals are likely to be included in the same community if the network around them is similar to that of their neighbors. The criteria for judging the efficacy of the partition of nodes into communities is embedded in the statistical model implied for the network and as such is a balance between all of the terms in the model. This contrasts a non-model-based objective function such as modularity which focuses on maximizing in some sense the ratio of density of ties within and between communities. To illustrate the difference, consider a $k$-star. The greater the value of $k$, the greater the discrepancy in the degree of the actors. Therefore, if $k$-stars occur frequently, the members of the same $k$-star are likely to be included in the same group by the latent class model but, due to the difference in degree, are unlikely to be grouped under modularity maximization. However, an advantage of the network science approach is that results are likely to be more robust to model mis-specifications than under the social network approach. 

In the future it is possible to imagine a bridging of the two approaches to community detection. For example, a model for the network or the component of the model involving the key determinants of network ties, could be incorporated in the modularity function in (\ref{eq:ModMax}). Depending on the specification, the result might be a weighted version of modularity in which a higher penalty is incurred if individuals with similar traits -- or in structurally equivalent positions with respect to $k$-stars, triadic closure or other local network configurations -- are included in different communities than if individuals with different traits are in different communities. However, to the best of the author's knowledge, such a procedure is not available.

\section*{Part IV: Discussion and Glossary}
\addtocounter{section}{1}
\label{sec:discuss}

In this chapter, the dual fields of social networks and network science have been described, with particular focus on sociocentric data. Both fields are growing rapidly in methodological results and the breadth of applications to which they are applied. 

In health applications, social network methods for evaluating whether individuals' attributes spread from person-to-person across a population (social influence) and for modeling relationship or tie status (social selection) have been described. Models of relationship status have not been applied as frequently in health applications, where focus often centers on the patient. However, \shortciteN{Keat:2007} is a notable exception. Due to the ever-growing availability of data, the interest in peer effects, and the need to design support mechanisms, the role of social network analysis in health care and medicine is likely to undergo continued growth in the future.

A novel feature of this chapter is the attention given to network science. Although network science is descriptively-inclined and thus is removed from mainstream translational medical research seeking to identify causes of medical outcomes, the increasing availability of complex systems data provides an opportunity for network science to play a more prominent role in medical research in the future. For example, Barabasi and others have created a Human Disease Network by connecting all hereditary diseases that share a disease-causing gene \shortcite{goh2007}. In other work, they created a Phenotypic Disease Network (PDN) as a map summarizing phenotypic connections between diseases \cite{Hida:2009}. These networks provided important insights into the potential common origins of different diseases, whether diseases progress through cellular functions (phenotypes) associated with a single diseased (mutated) gene or with other phenotypes, and whether patients affected by diseases that are connected to many other diseases tend to die sooner than those affected by less connected diseases. Such work has the potential to provide insights into many previously untested hypotheses about disease mechanisms. For example, they may ultimately be helpful in designing ``personalized treatments'' based on the network position held by an individual's combined genetic, proteomic, and phenotypic information. In addition, they may suggest conditions for which treatments found to be effective on another condition might also be tried.

There are several important topics that have not been discussed, notably including network sampling. In gathering network data, adaptive methods such as link-tracing designs are often used to identify individuals more likely to know each other and thus to have formed a relationship with other sampled individuals than in a random-probability design. Link-tracing and other related designs are often used to identify hard-to-reach populations \cite{Thom:1996,Thom:2000,Thom:2006}. However, the sampling probabilities corresponding to link-tracing designs may be difficult to evaluate (generally requiring the use of simulation) and it may not be obvious how they should be incorporated in the analysis. The development of statistical methods that account for the sample design in the analysis of social network data has lagged behind the designs themselves. However, recently progress has been made on statistical inference for sampled relational network data \shortcite{Hand:2010}. 

In the future it is likely that more bridges will form between the social network and the network science fields with models or methods developed in one field used to solve problems in the other. Furthermore, as these two fields become more entwined, it is likely that they will also become more prominent in the solution to important problems in medicine and health care. 

\section*{Acknowledgements}
The time and effort of Dr. O'Malley and Dr. Onnela on researching and developing this chapter was supported by NIH/NIA grant P01 AG031093 (PI: Christakis) and Robert Wood Johnson Award \#58729 (PI Christakis). Dr Onnela was further supported by NIH/NIAID grant  R01AI051164 (DeGruttola). The authors thank Mischa Haider, Brian Neelon, and Bruce E. Landon for reviewing an early draft of the manuscript and providing several useful comments and suggestions. 

\section*{Glossary of Terms}

To help readers familiar with social networks understand the network science component of the chapter and conversely for readers familiar with network science to understand the social network component, the following glossary contains a comprehensive list of terms and definitions. 

\subsection*{Terms used in social networks}

\begin{enumerate}
\item Social network: A collection of actors (referred to as actors) and the (social) relationships or ties linking them.
\item Relationship, Tie: A link or connection between two actors.
\item Dyad: A pair of actors in a network and the relationship(s) between them; two relationships per measure for a directed network, one relationship per measure for an undirected network.
\item Triad: A triple of three actors in the network and the relationships between them.
\item Scale or valued relationship: A non-binary relationship between two actors (e.g., the level of a trait). We focused on binary relationships in the chapter.
\item Directed network: A network in which the relationship from actor $i$ to actor $j$ need not be the same as that from actor $j$ to actor $i$.
\item Non-directed network: A network in which the state of the relationship from actor $i$ to actor $j$ equals the state of the relationship from actor $j$ to actor $i$.
\item Sociocentric network data: The complete set of observations on the $n(n-1)$ relationships in a directed network, or $n(n-1)/2$ relationships in an undirected network, with $n$ actors.
\item Collaboration network: A network whose ties represent the actors' joint involvement on a task (e.g., work on a paper) or a common experience (e.g., treating the same episode of health care for a patient). 
\item Bipartite: Relationships are only permitted between actors of two different types.
\item Unipartite: Relationships are permitted between all types of actors.
\item Social contagion, Social influence, Peer effects: Terms used to describe the phenomenon whereby an actor's trait changes due to their relationship with other actors and the traits of those actors.
\item Mutable trait: A characteristic of an actor than can change state.
\item Social selection: The phenomena whereby the relationship status between two actors depends on their characteristics, as occurs with homophily and heterophily.
\item Homophily: A preference for relationships with actors who have similiar characteristics. Popularly referred to as "birds of a feather flock together."
\item Heterophily: A preference for relationships with actors who have different characteristics. Popularly referred to as "opposites attracting."
\item In-degree, Popularity: The number of actors who initiated a tie with the given actor.
\item Out-degree, Expansiveness, Activity: The number of ties the given actor initiates with other actors.
\item $k$-star: A subnetwork in which the focal actor has ties to $k$ other actors.
\item $k$-cycle: A subnetwork in which each actor has degree 2 that can be arranged as a ring (i.e., a $k$-path through the actors returns to its origin without backtracking. For example, the ties A-B, B-C, and C-A form a three-cycle.
\item $k$ degrees of separation: Two individuals linked by a $k$-path ($k-1$ intermediary actors) that are not connected by any path of length $k-1$ or less.
\item Density: The overall tendency of ties to form in the network. A descriptive measure is given by the number of ties in the network divided by the total number of possible ties.
\item Reciprocity: The phenomena whereby an actor $i$ is more likely to have a tie with actor $j$ if actor $j$ has a tie with actor $i$. Only defined for directed networks.
\item Clustering: The tendency of ties to cluster and form densely connected regions of the network.
\item Closure: The tendency for network configurations to be closed.
\item Transitivity: The tendency for a tie from individual A to individual B to form if ties from individual A to individual C and from individual C to individual B exist. A form of triadic closure commonly stated as ``a friend of a friend is a friend.'' Reduces to general triadic closure in an undirected network.
\item Centrality: A dimenionless measure of an actors position in the network. Higher values indicate more central positions. There are numerous measures of centrality. Four common ones are degree, closeness, betweeness, and eigenvalue centrality. Degree and eigenvalue centrality are extremes in that degree centrality  is determined solely from an actor's degree (it is internally focused) while eigenvalue centrality is based on the centrality of the actors connected to the focal actor (it is externally focused).
\item Structural balance: A theory which suggests actors seek balance in their relationships; for example, if A likes B and B likes C then A will endeavor to like C as well to keep the system balanced. Thus, the existence of transitivity is implied by structural balance.
\item Structural equivalence: The network configuration (arrangement of ties) around one actor is similar to that of another actor. Even though actors may not be connected, they can still be in structurally similar situations.
\item Structural power: An actor in a dominant position in the network. Such an actor may be one in a strategic position, such as the only bridge between otherwise distinct components. 
\item Network component: A subset of actors having no ties external to themselves.
\item Graph theory: The mathematical basis under which theoretical results for networks are derived and empirical computations are performed.
\item Digraph: A graph in which edges can be bidirectional. Unlike social networks, digraphs can contain self-ties. Graphs lie in two-dimensional space.
\item Hypergraph: A graph in dimension three or higher.
\item Maximal subset: A set of actors for whom all ties are intact in a binary-network (i.e., has density 1.0). If the set contains $k$ actors, the maximal subset is referred to as a $k$-clique.
\item Scalar, vector, matrix: Terms from linear and abstract algebra. A scalar is a $1 \times 1$ matrix, a vector is a $k \times 1$ matrix, and a matrix is $k \times p$, where $k, p > 1$.
\item Adjacency matrix: A matrix whose off-diagonal elements contain the value of the relationship from one actor to another. For example, element $ij$ contains the relationship from actor $i$ to actor $j$. The diagonal elements are zero by definition.
\item Matrix transpose: The operation whereby element $ij$ is exchanged with element $ji$ for all $i,j$.
\item Row stochastic matrix: A matrix whose rows sum to 1 and contain non-negative elements. Thus, each row represents a probability distribution of a discrete-valued random variable. 
\item Random variable: A variable whose value is not known with certainty. It can relate to an event or time period that is yet to occur, or it can be a quantity whose value is fixed (i.e., has occurred) but is unknown.
\item Parametric: A term used in statistics to describe a model with a specific functional form (e.g., linear, quadratic, logarithmic, exponential) indexed by unknown parameters or an estimation procedure that relies on specification of the complete distribution of the data.
\item Non-parametric: A model or estimation procedure that makes no assumption about the specific form of the relationship between key variables (e.g., whether the predictors have linear or additivie effects on the outcome) and does not rely upon complete specification of the distribution of the data for estimation.
\item Outcome, Dependent variable: The variable considered causally dependent on other variables of interest. This will typically be a variable whose value is believed to be caused by other variables.
\item Independent, Predictor, Explanatory variable, Covariate: A variable believed to be a cause of the outcome.
\item Contextual variable: A variable evaluated on the neighbors of, or other members of a set containing, the focal actor. For example, the proportion of females in a neighboring county, the proportion of friends with college degrees.  
\item Interaction effect: The extent to which the effect of one variable on the outcome varies across the levels of another variable.
\item Endogenous variable: A variable (or an effect) that is internal to a system. Predictors in a regression model that are correlated with the unobserved error are endogeneous; they are determined by an internal as opposed to an external process. By definition outcome variables are endogenous.
\item Exogenous variable: A variable (or an effect) that is external to the system in that its value is not determined by other variables in the system. Predictors that are independent of the error term in a regression model are exogeneous.
\item Instrumental variable (IV): A variable with a non-null effect on the endogeneous predictor whose causal effect is of interest (the "treatment") that has no effect on the outcome other than that through its effect on treatment. Often-used sufficient conditions for the latter are that the IV is (i) marginally independent of any unmeasured confounders and (ii) conditionally independent of the outcome given the treatment and any unmeasured confounders. In an IV analysis a set of observed predictors may be conditioned on as long as they are not effects of the treatment and the IV assumptions hold conditional on them. While subject to controversy, IV methods are one of the only methods of estimating the true (causal) effect of an endogeneous predictor on an outcome.
\item Linear regression model: A model in which the expected value of the outcome (or dependent variable) conditional on one or more predictors (or explanatory variables) is a linear combination of the predictors (an additive sum of the predictors multiplied by their regression coefficients) and an unobserved random error.
\item Longitudinal model: A model that describes variation in the outcome variable over time as a function of the predictors, which may include prior (i.e., lagged) values of the outcome. Observations are typically only available at specific, but not necessarily equally-spaced, times. Longitudinal models make the direction of causality explicit. Therefore, they can distinguish between the association between the predictors and the outcome and the effect of a change in the predictor on the change in the outcome.
\item Cross-sectional model: A model of the relationship between the values of the predictors and outcomes at a given time. Because one cannot discern the direction of causality, cross-sectional models are more difficult to defend as causal.
\item Stochastic block model: A conditional dyadic independence model in which the density and reciprocity effects differ between blocks defined by attributes of the actors comprising the network. For example, blocks for gender accomodate different levels of connectedness and reciprocity for men and women.
\item Logistic regression: A member of the exponential family of models that is specific to binary outcomes. It utilizes a link function that maps expected values of the outcome onto an unrestricted scale to ensure that all predictions from the model are well-defined.
\item Multinomial distribution: A generalization of the binomial distribution to three or more categories. The sum of the probabilities of each category equals 1.
\item Exponential random graph model: A model in which the state of the entire network is the dependent variable. Provides a flexible approach to accounting for various forms of dependence in the network. Not amenable to causal modeling.
\item Degeneracy: An estimation problem encountered with exponential random graph models in which the fitted model might reproduce observed features of the network on average but each actor draw bears no resemblence to the observed network. Often degenerate draws are empty or complete graphs.
\item Latent distance model: A model in which the status of dyads are independent conditional on the positions of the actors, and thus the distance between them, in a latent social space.  
\item Latent eigenmodel: A model in which the status of dyads are independent conditional on the product of the (weighted) latent positions of the actors in the dyad. 
\item Latent variable: An unobserved random variable. Random effects and pure error terms are latent variables.
\item Latent class: An unobserved categorical random variable. actors with the same value of the variable are considered to be in the same latent class.
\item Factor analysis: A statistical technique used to decompose the correlation (or covariance) matrix of a set of random variables into groups of related items. 
\item Generalized estimating equation (GEE): A statistical method that corrects estimation errors for dependent observations without necessarily modeling the form of the dependence or specifying the full distribution of the data.
\item Random effect: A parameter for the effect of a unit (or cluster) that is drawn from a specified probability distribution. Treating the unit effects as random draws from a common probability distribution allows information to be pooled across units for the estimation of each unit-specific parameter.
\item Fixed effect: A parameter in a model that reflects the effect of an actor belonging to a given unit (or cluster). By virtue of modeling the unit effects as unrelated parameters, no information is shared between units and so estimates are based only on information within the unit.
\item Ordinary least squares: A commonly-used method for estimating the parameters of a regression model. The objective function is to minimize the squared distance of the fitted model to the observed values of the dependent variable.
\item Maximum likelihood: A method of estimating the parameters of a statistical model that typically embodies parametric assumptions. The procedure is to seek the values of the parameters that maximize the likelihood function of the data. 
\item Likelihood function: An expression that quantifies the total information in the data as a function of model parameters. 
\item Markov chain Monte Carlo: A numerical procedure used to fit Bayesian statistical models.
\item Steady state: The state-space distribution of a Markov chain describes the long-run proportion of time the random variable being modeled is in each state. Often Markov chains iterate through a transient phase in which the current state of the chain depends less and less on the initial state of the chain. The steady state phase occurs when successive samples have the same distribution (i.e., there is no dependence on the initial state).
\item Colinearity: The correlation between two predictors after conditioning on the other observed predictors (if any). When predictors are colinear distinguishing their effects is difficult and the statistical properties of the estimated effects are more sensitive to the validity of the model.
\item Normal distribution: Another name for the Gaussian distribution. Has a bell-shaped probability density function.
\item Covariance matrix: A matrix in which the $ij$th element contains the covariance of items $i$ and $j$.
\item Absolute or Geodesic distance: The total distance along the edges of the network from one actor to another.
\item Cartesian distance: The distance between two points on a two-dimension surface or grid. Adheres to Pythagorus Theorem.
\item Count data: Observations made on a variable with the whole numbers (0, 1, 2, $\ldots$) as its state space.
\item Statistical inference: The process of establishing the level of certainty of knowledge about unknown parameters (or hypothesis) from data subject to random variation, such as when observations are measured imperfectly with no systematic bias or a sample from a population of interest is used to estimate population parameters.
\item Null model: The model of a network statistic typically represents what would be expected if the feature of interest was non-existent (effect equal to 0) or outside the range of interest. 
\item Permutation test: A statistical test of a null hypothesis against an alternative implemented by randomly re-shuffling the labels (i.e., the subscripts) of the observations. The significance level of the test is evaluated by re-sampling the observed data 50 -- 100 times and computing the proportion of times that the test is rejected.
\end{enumerate}

\subsection*{Terms used in network science}

\begin{enumerate}
\item Network science: The approach developed from 1995 onwards mostly within statistical physics and applied mathematics to study networked systems across many domains (e.g., physical, biological, social, etc). Usually focuses on very large systems; hence, theoretical results derived in the thermodynamic limit are good approximations to real-world systems.
\item Thermodynamic limit: In statistical physics refers to the limit obtained for any quantity of interest as system size $N$ tends to infinity. Many analytical results within network science are derived in this limit due to analytical tractability.
\item Statistical physics: The branch of physics dealing with many body systems where the particles in the system obey a fix set of rules, such as Newtonian mechanics, quantum mechanics, or any other rule set. As the number of bodies (particles) in a system grows, it becomes increasingly difficult (and less informative) to write down the equations of motion, a set of differential equations that govern the motion of the particles over time, for the system. However, one can describe these systems probabilistically. The word ``statistical'' is somewhat misleading as there is no statistics in the sense of statistical inference involved; instead everything proceeds from a set of axioms, suggesting that ``probabilistic'' might be a better term. Statistical physics, also called statistical mechanics, gives a microscopic explanation to the phenomena that thermodynamics explains phenomenologically.
\item Generative model: Most network models within network science belong to this category. Here one specifies the microscopic rules governing, for example, the attachment of new nodes to the existing network structure in models of network growth.
\item Cumulative advantage: A stylized modeling mechanism introduced by Price in 1976 to capture phenomena where ``success breeds success.'' Price applied the model to study citation patterns where power-law or power-law like distributions are observed for the distribution of the number of citations and successfully reproduced by the model.
\item Polya urn model: A stylized sampling model in probability theory where the composition of the system, the contents of the urn, changes as a consequence of each draw from the urn.
\item Power law: Refers to the specific functional form $P(x) \sim x^{-\alpha}$ of the distribution of quantity $x$. Also called Pareto distribution. See scale-free network.
\item Preferential attachment: A stylized modeling mechanism introduced by Barabasi and Albert in 1999 where the probability of a new node to attach itself to an existing node $i$ of degree $k_i$ is an increasing function of $k_i$; in the case of linear preferential attachment, this probability is directly proportional to $k_i$. In short, the higher the degree of a node, the higher the rate at which it acquires new connections (increases its degree).
\item Weak ties hypothesis: A hypothesis developed by sociologist Mark Granovetter in his extremely influential 1973 paper ``The strength of weak ties.'' The hypothesis, in short, states the following: The stronger the tie connecting persons $A$ and $B$, the higher the fraction of friends they have in common.
\item Modularity: Modularity is a quality-function used in network community detection, where its value is maximized (in principle) over the set of all possible partitions of the network nodes into communities. Standard modularity reads as $Q=(2m)^{-1} \sum_{i,j} (A_{ij} - \frac{k_i k_j}{2m}) \delta(c_i, c_j)$ where $c_i$ is the community assignment of node $i$ and $\delta$ is Kronecker delta; other quantities as defined in the text.
\item Rate equations: Rate equations, commonly used to model chemical reactions, are similar to master equations but instead of modeling the count of objects (e.g., number of nodes) in a collection of discrete states (e.g., the number of $k$-degree nodes $N_k(t)$ for different values of $k$), they are used to model the evolution of continuous variables, such as average degree, over time.
\item Master equations: Widely used in statistical physics, these differential equations model how the state of the system changes from one time point to the next. For example, if $N_k(t)$ denotes the number of nodes of degree $k$, given the model, one can write down the equation for $N_k(t+1)$, i.e., the number of $k$-degree nodes at time $t+1$.
\item Fitness or affinity or attractiveness:  A node attribute introduced to incorporate heterogeneity in the node population in a growing network model. For example, in a model based on preferential attachment, this could represent the inherent ability of a node to attract new edges, a mechanism that is superimposed on standard preferential attachment.
\item Community: A group of nodes in a network that are, in some sense, densely connected to other nodes in the community but sparsely connected to nodes outside the community.
\item Community detection: The set of methods and techniques developed fairly recently for finding communities in a given network (graph). The number of communities is usually not specified {\it a priori} but, instead, needs to be determined from data.
\item Critical point: The value of a control parameter in a statistical mechanical system where the system exhibits critical behavior: previously localized phenomena now become correlated throughout the system which at this point behaves as one single entity.
\item Phase diagram: A diagram displaying the phase (liquid, gas, etc.) of the system as one or more thermodynamic control parameters (temperature, pressure, etc.) are varied.
\item Phase transition: Thermodynamic properties of a system are continuous functions of the thermodynamic parameters within a phase; phase transitions (e.g., liquid to gas) happen between phases where thermodynamic functions are discontinuous.
\item Network diameter: The longest of the shortest pairwise paths in the network, computed for each dyad (node pair).
\item Hysteresis: The behavior of a system depends not only on its current state but also on its previous state or states.
\item Quality function: Typically a real-valued function with a high-dimensional domain that specifies the ``goodness'' of, say, a given network partitioning. For example, given the community assignments of $N$ nodes, which can be seen as a point in an $N$-dimensional hypercube, the standard modularity quality function returns a number indicating how good the given partitioning is.
\item Dynamic process: Any process that unfolds on a network over time according to a set of pre-specified rules, such as epidemic processes, percolation, diffusion, synchronization, etc.
\item Slice: In the context of multi-slice community detection, refers to one graph in a collection of many within the same system, where a slice can capture the structure of a network at a given time (time-dependent slice), at a particular resolution level (multiscale slice), or can encode the structure of a network for one tie type when many are present (multiplex slice).
\item Scale-free network: Network with a power-law (Pareto) degree distribution.
\item \er model: Also known as Poisson random graph (after the fact that the degree distribution in the model follows a Poisson distribution), Bernoulli random graph (after the fact that each edge corresponds to an outcome of a Bernoulli process), or the random graph (as the progenitor of all random graphs). Starting with a fixed set of $N$ nodes, one considers each node pair in turn independently of the other node pairs and connects the nodes with probability $p$. Erd\H{o}s and R\'{e}nyi first published the model in 1959, although Solomonoff and Rapoport published a similar model earlier in 1951.
\item \ws model: A now canonical model by Watts and Strogatz that was introduced in 1998. Starting from a regular lattice structure characterized by high clustering and long paths, the model shows how randomly rewiring only a small fraction of edges (or, alternative, adding a small number of randomly placed edges) leads to a small-world characterized by high clustering and short paths. The model is conceptually appealing, and shows how to interpolate, using just one parameter, from a regular lattice structure in one extreme to an \er graph in the other.
\item Mean-field approximation: Sometimes called the zero-order approximation, this approximation replaces the value of a random variable by its average, thus ignoring any fluctuations (deviations) from the average that may actually occur. This approach is commonly used in statistical physics.
\item Ensemble: A collection of objects, such as networks, that have been generated with the same set of rules, where each object in the ensemble has a certain probability associated with it. For example, one could consider the ensemble of networks that consists of 6 nodes and 2 edges, each begin equiprobable.
\end{enumerate}

\end{document}